\documentclass{article}
\usepackage{authblk}
\usepackage{dsfont}  
\usepackage{pdfpages}
\usepackage{pgf}

\usepackage[letterpaper]{geometry}
\usepackage{natbib}
\usepackage{graphicx}
\usepackage{multirow}
\usepackage[centertags]{amsmath}
\usepackage{rotating}
\usepackage{multicol}        
\usepackage{multicol}
\usepackage{times}

\usepackage{url}

\def\R{\mathds{R}}
\def\N{\mathds{N}}
\def\Q{\mathcal{Q}}
\def\S{\mathcal{S}}

\def\x{{\bf x}}
\def\X{{\bf X}}

\def\q{{\bf q}}
\def\btheta{{\boldsymbol\theta}}
\def\blambda{{\boldsymbol\lambda}}
\def\brho{{\boldsymbol\rho}}
\def\bxi{{\boldsymbol\xi}}

\def\argmin{\mathop{\rm arg\,min}\limits}

\newtheorem{theorem}{Theorem} 

\newtheorem{definition}{Definition} 
\newtheorem{proposition}{Proposition} 
\newcommand{\indicator}[1]{\mathds{1}_{\left[ {#1} \right] }}

\title{Quantile-based clustering}
\author[1]{Christian Hennig\footnote{mailto: \texttt{christian.hennig@unibo.it}}}
\author[1]{Cinzia Viroli}
\author[1]{Laura Anderlucci}
\affil[1]{Department of Statistical Sciences, University of Bologna, Italy.}

\date{}

\begin{document}

\maketitle

\begin{abstract} 
A new cluster analysis method,
$K$-quantiles clustering, is introduced. $K$-quantiles clustering can be
computed by a simple
greedy algorithm in the style of the classical Lloyd's algorithm
for $K$-means. It can be applied to large and high-dimensional datasets.
It allows for within-cluster skewness and internal variable scaling based on
within-cluster variation. Different versions allow for different levels of
parsimony and computational efficiency. Although $K$-quantiles clustering
is conceived as nonparametric, it can be connected to a fixed partition model
of generalized asymmetric Laplace-distributions. The consistency of
$K$-quantiles clustering is proved, and it is shown that $K$-quantiles
clusters correspond to well separated mixture components in a nonparametric
mixture. In a simulation, $K$-quantiles clustering is compared with a number
of popular clustering methods with good results. A high-dimensional
microarray dataset is clustered by $K$-quantiles.

\end{abstract}


\section{Introduction}

In this paper we introduce a new clustering method, quantile-based or
$K$-quantiles clustering.
The method is fast and simple and can deal with large datasets. A special
feature of the method is that it takes into account potential skewness of
the within-cluster distributions.

The popular $K$-means method \citep{Jain10} represents all clusters by their
centroids (cluster means) and assigns all points to the closest centroid.
Quantile-based clustering represents the clusters by optimally chosen
quantiles. Points are assigned to the closest quantile (or rather, in
multidimensional data, distances to quantiles are summed up over the
variables), but the distance measuring ``closeness'' treats points
asymmetrically depending on which side of the quantile they are. This idea
has been explored for supervised classification by \cite{HeVi16}, and here we
present its application to clustering.

The algorithm for $K$-quantiles clustering works along
the lines of Lloyd's
classical $K$-means algorithm \citep{Lloyd82}
and is in this way faster and simpler than
many modern clustering methods, at the same time being more flexible than
$K$-means.

There is some ambiguity in the literature about to what extent $K$-means is
model-based. The $K$-means objective function can be motivated without
reference to probability models; it formalizes optimal representation of all
points in a cluster by the cluster centroid in the sense of least squares. It
is therefore sometimes presented as assumption-free method. But $K$-means can
also be derived as Maximum Likelihood (ML) estimator of a fixed partition model
of spherical Gaussian clusters with equal within-cluster variances, which seems
to be a quite severe assumption. Indeed $K$-means tends to produce spherical
clusters, so it is hardly appropriate to call it assumption-free, although it
can be applied to data that do not follow this model assumption. Whether this
is appropriate does not depend so much on to what extent the model assumption
is really fulfilled, but rather on whether the $K$-means characteristics matches
the ``shape'' of clusters required in the application in hand. Different
applications of cluster analysis ask for different kinds of clusters, and the
user of cluster analysis needs to understand such characteristics of methods
in order to choose an appropriate one for the application of interest
\citep{Hennig16}.

In the same way, $K$-quantiles clustering can also be derived as ML-estimator
for a fixed partition model of generalized asymmetric Laplace distributions.
This
is helpful also for the construction of $K$-quantiles clustering, because it
implies how to penalize variables against each other when using different
quantiles for different variables. It also allows for an in-built scaling of
variables that takes skewness into account. However, the main rationale of
$K$-quantiles clustering is not the estimation of asymmetric
Laplace distributions,
but rather to define a general clustering principle that is almost as simple as
$K$-means but more flexible by taking within-cluster skewness into account.
Throughout the paper, the number of clusters $K$ is treated as fixed; the estimation of $K$ is left to future work.

We review the principle of $K$-means clustering in Section \ref{skmeans}. In
Section \ref{sqcluster}, quantile-based clustering is motivated and defined.
First, we motivate it in a discrepancy-based nonparametric manner. Then we
link it to a fixed partition model of asymmetric Laplace distributions. Some
attention is paid to the penalty term introduced by ML-estimation in this model.
A simple greedy algorithm is proposed, and various constrained versions of the
quantile-based clustering are proposed, which allow for more parsimony and
less computational effort. Section \ref{sconsistency} is devoted to
consistency theory. Quantile-based clustering is proved to be consistent in a
nonparametric setting for the canonical clustering functional defined on a
distribution, and another theorem shows that this
functional will yield clusters that correspond to mixture components in mixtures
with strongly separated nonparametric components. Section \ref{sec:sims} presents a simulation study that includes high-dimensional setups, in which $K$-quantiles clustering is compared with some popular clustering methods. In Section
\ref{sgene}, $K$-quantiles clustering is applied to a real microarray dataset
with more than 3000 variables. Section \ref{sconc} concludes the paper.

\section{$K$-means and distance-based probabilistic clustering}\label{skmeans}
The aim of fixed partition clustering is to seek the best partition of $n$ data vectors $\tilde \x_n=(\x_1,\ldots,\x_n) \in (\R^p)^n$
into $K$ disjoint subsets characterized by cluster prototypes $\tilde\bxi=(\boldsymbol\xi_1,\ldots,\boldsymbol\xi_K)$ (the tilde denotes a collection of vectors rather than a single one).

In classical $K$-means \citep{Jain10} the `best' partition $C=(C(1),\ldots,C(n))$ is obtained by minimizing over $\tilde\bxi$ and $C$ the variance function given by
\begin{eqnarray}\label{eqn:kmeans}
V^{K-means}_{n,K}(\tilde\bxi,C,\tilde\x_n)=\sum_{i=1}^n \| \textbf{x}_i - \boldsymbol\xi_{C(i)}\|^2,
\end{eqnarray}
where $\| \bullet \|$ denotes the $L_2$ or Euclidean distance,
$C(i)\in\{1,\ldots,K\}$ for $i=1,\ldots,n$.
A classical estimation algorithm for minimizing $V^{K-means}_{n,K}$ consists of two steps sequentially iterated until convergence \citep{Lloyd82}. In the first step, for fixed $\boldsymbol\xi$ the best partition $C$ is found by assigning each point to the nearest cluster center. Then in the second step, for fixed $C$, the centroids $\boldsymbol\xi_k$ ($k=1,\ldots,K$) are estimated. Since the sum of squared Euclidean distances in (\ref{eqn:kmeans}) is minimized by the mean, the centroids $\boldsymbol\xi_k$ ($k=1,\ldots,K$) are the within cluster means.

Although usually no probability assumption is mentioned when $K$-means is
introduced, $K$-means can
be derived as Maximum Likelihood (ML) estimator of a fixed partition model
of spherical Gaussian clusters with equal within-cluster variances. According to such a model, $\x_1,\ldots,\x_n$ are independently drawn from
${\cal N}(\boldsymbol\xi_{C(i)},\sigma^2 \textbf{I}_p),\ i=1,\ldots,n$, where $C(i)\in\{K=1,\ldots,K\}$ are parameters giving the cluster memberships of the $\x_i$; as opposed to a mixture model, in a fixed partition model these are not modelled as random. The log-likelihood of such a model is
\[
-\sum_{i=1}^n \frac{p}{2}\log \sigma^2 -\left\{\frac{1}{2\sigma^2} \| \textbf{x}_i - \boldsymbol\xi_{C(i)}\|^2 \right\},
\]
which is maximized by the $\bxi_1,\ldots,\bxi_K,C(1),\ldots,C(n)$ that minimize
$V^{K-means}_{n,K}$, in other words, by $K$-means.

More generally, starting from an arbitrary distance from a prototype, denoted by $d(x,\xi)$, it is always possible to construct a probabilistic clustering model as proposed by \cite{israel} and \cite{Iyigun}. The kernel of the distance-based density is the inverse of the exponential of the distance measure weighted by a positive concentration parameter $\lambda$:
\begin{eqnarray}\label{eqn:distance}
  f(x;\xi,\lambda)=\psi(\xi,\lambda)e^{-\lambda d(x,\xi)}
  \end{eqnarray}
where $d(x,\xi)$ is a generic distance function from a location parameter $\xi$, $\lambda > 0$, and $\psi(\xi,\lambda)$ is a normalization constant such that $f(\textbf{x};\xi,\lambda)$ is a proper density function.

Distance-based models have been used by several authors \citep[see][]{Ma57,FV86,Di88} and adapted for classification in a mixture-based perspective by \cite{MM03} for ranking data and by \cite{banerjee} for textual data.

Note that, when $d(x,\xi)$ is the $L_2$ (Euclidean) distance from the expected value of $X$, $\xi=E[X]$, the density (\ref{eqn:distance}) is the Gaussian distribution. When $d(x,\xi)$ is the $L_1$ distance, the density (\ref{eqn:distance}) coincides with the Laplace distribution. When $d(x,\xi)$ is the cosine distance and data are normalized to 1 according to the $L_2$ norm, (\ref{eqn:distance}) becomes the von Mises-Fisher distribution \citep{banerjee}.

\section{Quantile-based clustering}\label{sqcluster}
We now introduce a new clustering strategy based on the idea of assigning points to the closest quantile. Measuring ``closeness'' by the squared Euclidean distance is associated with the mean, in the sense that means optimize (\ref{eqn:kmeans}). Quantiles can also be characterized by minimizing a sum of discrepancies, although these discrepancies are not symmetric; they depend on which side of the quantile a point is. Using these discrepancies in ``$K$-means style'' leads to a simple clustering method that allows for within-cluster skewness.

\subsection{Clustering based on the quantile discrepancy}
Let $X$ be a univariate random variable defined on $\R$ with probability cumulative function $F_X(x)$. Let $\theta \in [0,1]$ be a percentile and denote as $q(\theta)$ the corresponding quantile, such as $F_X^{-1}(\theta)=q(\theta)=\inf\{x:F_X(x)\geq \theta\}$.

The quantile $q(\theta)$ is the not necessarily unique
value of $\xi$ that minimizes
the following variability measure:
\begin{eqnarray}\label{eqn:qdist.centroid.0}
\theta\int_{x>\xi}|x-\xi|dF_X(x)+(1-\theta)\int_{x< \xi}|x-\xi|dF_X(x).
\end{eqnarray}
For a single point $x$, we define the quantile discrepancy from $\xi$ as a function $\Q: \ \R \times [0,1] \rightarrow [0,\infty)$:
\begin{eqnarray}\label{eqn:qdist.centroid}
\Q(x, \theta,\xi)=\left\{\theta+(1-2\theta)\indicator{x<\xi} \right \}|x-\xi|.
\end{eqnarray}
For $\theta=0.5$, this is the $L_1$ distance, but for $\theta\neq 0.5$ it is not
symmetric and therefore not a distance. Not being based on squares, it shares with the $L_1$ distance its better resistance against outliers compared to the $L_2$ distance.




By definition the quantile discrepancy has a univariate nature. When $X$ is a multivariate random variable on $\R^p$, the quantile discrepancy with respect to a generic vector of centroids $\boldsymbol\xi$ is defined as the sum of component-wise distances:
\begin{equation}\label{eqn:qdistp.centroid}
\Q^*(\x,\boldsymbol\theta,\boldsymbol\xi)=\sum_{j=1}^p\Q(x_j,\theta_j,\xi_j)=\sum_{j=1}^p\left\{\theta_j+(1-2\theta_j)\indicator{x_j<\xi_j} \right \}|x_j-\xi_j|,
\end{equation}
where $\theta_j$ can be variable-wise or a single common percentile for all variables.

The basic idea of quantile-based clustering is to use the quantile discrepancy instead of the squared $L_2$-distance in $K$-means, i.e., minimizing
\begin{equation}\label{eq:quantilenopen}
V^{K-quantiles}_{n,K}(\btheta,\tilde\bxi,C,\tilde\x_n)=\sum_{i=1}^n
\sum_{j=1}^p
\Q(x_{ij},\theta_j,\xi_{C(i)j}),
\end{equation}
where again $\tilde\bxi=(\bxi_1,\ldots,\bxi_K)$.
$\btheta$ is assumed here to be the same for all clusters. $\bxi_1,\ldots,\bxi_K$ define the locations of the clusters. We call them ``barycenters'' from now on.

\begin{proposition} \label{pquantile}
Let $\x_1,\ldots,\x_n\in \R^p$, $\boldsymbol\theta\in (0,1)^p$ and $C(1),\ldots,C(n)\in\{1,\ldots,K\}$ so that $n_k=|\{\x_i:\ C(i)=k\}|>0$.
Then the empirical quantile vectors
$\q_{nk}(\boldsymbol\theta)=\{q_{nk1}(\theta_1),\ldots,q_{nkp}(\theta_p)\}$,
$k=1,\ldots,K$,
defined for $j=1,\ldots,p$ as
\[
q_{nkj}(\theta_j)=\inf\left\{x_j:\ \frac{1}{n_k}\sum_{C(i)=k} \indicator{x_{ij}\leq x_j} \geq \theta_j\right\}
\]
satisfy
\begin{equation}\label{eqn:prop1}
\sum_{i=1}^n  \sum_{j=1}^p
\Q(x_{ij},\theta_j,q_{nkj}(\theta_j))=\min_{\tilde\bxi}\sum_{i=1}^n
\sum_{j=1}^p \Q(x_{ij},\theta_j,\xi_{C(i)j}).
\end{equation}
\end{proposition}

\medskip
The proof of Proposition \ref{pquantile} is given in Appendix \ref{proof:pquantile}.
%

\medskip
(\ref{eq:quantilenopen}) quantifies the discrepancy between the observations in a cluster and their barycenter given $\btheta$, and is therefore appropriate
for finding the cluster barycenters and clustering the points,
but it will not work well for finding $\btheta$. The problem of finding
the optimal $\btheta$ will benefit from a model-based approach.

%
%



\subsection{The fixed partition model and quantile-based clustering} \label{sfp}
Quantile-based clustering can be derived as ML-estimator of a probabilistic model, similarly to $K$-means.

Consider the quantile discrepancy at $\xi=q(\theta)$, inserting $d(x,\xi)=\Q(x,q(\theta),\theta)$ in the distance-based density in (\ref{eqn:distance}) with $p=1$ for the moment.
The normalization constant is dependent on $\theta$,
and on $\xi$ only through $\theta$, so we can write
$\psi(\theta,\xi,\lambda)=\psi(\theta,\lambda)=\lambda\theta(1-\theta)$. Therefore the quantile discrepancy based density is a spiky curve taking the general form:
\begin{eqnarray}\label{eqn:qdist2}
f(x;\theta,\xi,\lambda)=\lambda\theta(1-\theta)e^{-\lambda \left\{\theta+(1-2\theta)\indicator{x<\xi} \right \}|x-\xi|},
\end{eqnarray}
where $\xi=q(\theta)$.

\begin{figure}
  \centering
\makebox{\includegraphics[width=\textwidth]{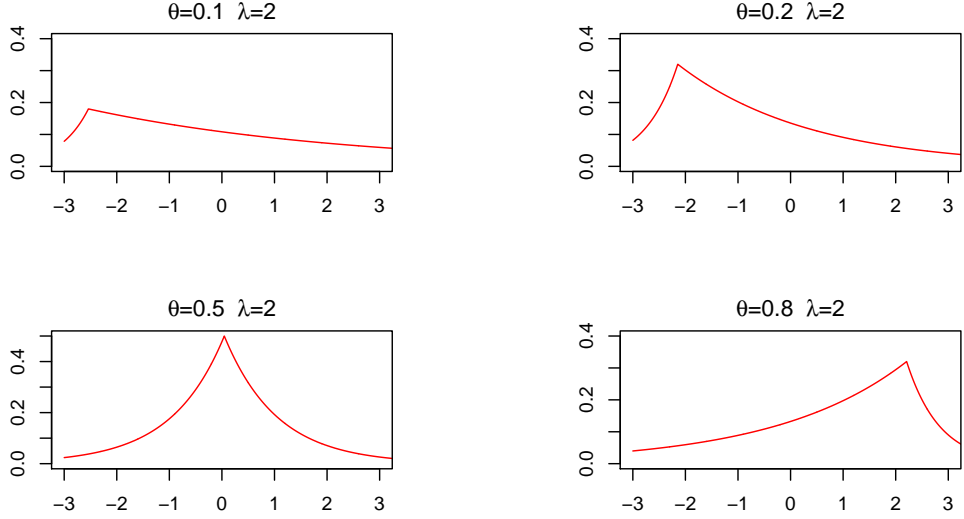}}
  \caption{\label{fig: dynamic}Examples of quantile-based densities for different values of $\theta$.}
\end{figure}

When $\theta=0.5$, then $\xi=q(1/2)$ is the median, and the quantile-based density is the Laplace distribution. When $\theta \neq 0.5$ the quantile-based density is a special case of the asymmetric Laplace distribution \citep{ald}
with expectation $E[X]=\xi + \frac{1-2\theta}{\lambda\theta(1-\theta)}$, variance $Var[X]=\frac{1-2\theta(1-\theta)}{(\lambda\theta(1-\theta))^2}$ and skewness $Skew[X]=\frac{(2 (1 - 2 \theta) (1 - (1 - \theta) \theta))}{(1 - 2 (1 - \theta) \theta)^{3/2}}$. Figure \ref{fig: dynamic} shows some examples of its shape as $\theta$ varies.

For $\tilde \x_n=(\x_1,\ldots,\x_n) \in (\R^p)^n$
we assume that the $p$ variables are independent
within clusters, and that the parameters $\theta$ and $\lambda$ do not differ between clusters; the clusters are distinguished only by different barycenters $\bxi_k,\ k=1,\ldots,K$. We aim at finding a compromise here between flexibility on one side and parsimony and computational simplicity on the other side. In the $K$-means model, variables are independent, all variables have the same within-cluster variances and clusters only differ regarding their centers. For quantile-based clustering, we define different levels of flexibility, see Section \ref{sgen}. For the moment we focus on the most general case of the models considered there, which allows both $\theta$ and $\lambda$ to vary between variables, allowing for different distributional shapes and scales. Allowing them to differ
between clusters as well, and incorporating within-cluster
dependence would define a considerably more complex approach, both regarding the number of parameters and the computational burden. This is left for future research.

With parameter vector
$\Theta=(\btheta, \tilde\bxi, \blambda,C)$,
the likelihood for a fixed partition asymmetric Laplace distribution model
with independent variables is
\[
f(\tilde\x_n;\Theta)=\prod_{i=1}^n \prod_{j=1}^p f(x_{ij};\theta_j,\xi_{C(i)j},\lambda_j),
\]
Plugging in (\ref{eqn:qdist2}) and taking logs, the ML estimator is
\begin{eqnarray}
T_{n,K}(\tilde\x_n) &=& \arg\min_{\Theta} V_{n,K}(\Theta,\tilde\x_n), \nonumber\\
V_{n,K}(\Theta,\tilde\x_n) &=& \sum_{i=1}^n
\sum_{j=1}^p \lambda_j \Q(x_{ij},\theta_j,\xi_{C(i)j}) -n \sum_{j=1}^p \log \lambda_j\theta_j(1-\theta_j), \label{eq:defv}
\end{eqnarray}
which for given $\btheta$ and $\blambda={\bf 1}$
leads to the same clustering as (\ref{eq:quantilenopen}).
Proposition \ref{pquantile} enforces that $\xi_{11},\ldots,\xi_{Kp}$ are the
variable-wise within-cluster $\theta$-quantiles, because the minimization
with respect to $\tilde\bxi$ is independent of $\blambda$.
For given $(\btheta, \tilde\bxi, \blambda)$, the ML-estimator of the clustering
$C$ is, for $i=1,\ldots,n$:
\begin{equation}\label{eq:clustering}
C(i)=\argmin_{k\in\{1,\ldots,K\}} \sum_{j=1}^p \lambda_j \Q(x_{ij},\theta_j,\xi_{kj}) -n \sum_{j=1}^p \log \lambda_j\theta_j(1-\theta_j).
\end{equation}
We will therefore omit $C$ in the parameter vector in the following. Here is
the resulting definition.

\begin{definition} \label{dkquantiles}
Quantile-based ($K$-quantiles) clustering
(with variable-wise $\theta$ and $\lambda$)
is defined by
\begin{equation}\label{eq:estimation}
T_{n,K}(\tilde\x_n)= \argmin_{\btheta,\tilde\bxi,\blambda}
\sum_{i=1}^n \min_{k\in\{1,\ldots,K\}}
\sum_{j=1}^p \lambda_j \Q(x_{ij},\theta_j,\xi_{kj}) -n \sum_{j=1}^p \log \lambda_j\theta_j(1-\theta_j);
\end{equation}
observations are clustered by (\ref{eq:clustering}) based on
$(\btheta,\tilde\bxi,\blambda)=T_{n,K}(\tilde\x_n)$.
\end{definition}

\subsection{Notes on penalization and scaling}\label{spenscale}

Comparing (\ref{eq:quantilenopen}) and (\ref{eq:estimation}) shows that
the logarithmized normalization constant
$-n \sum_{j=1}^p \log \lambda_j\theta_j(1-\theta_j)$ acts as a penalty term,
penalizing $\theta_j$ too close to 0 or 1 and too small $\lambda_j$.

\begin{figure}
  \centering
   \makebox{\includegraphics[width=\textwidth]{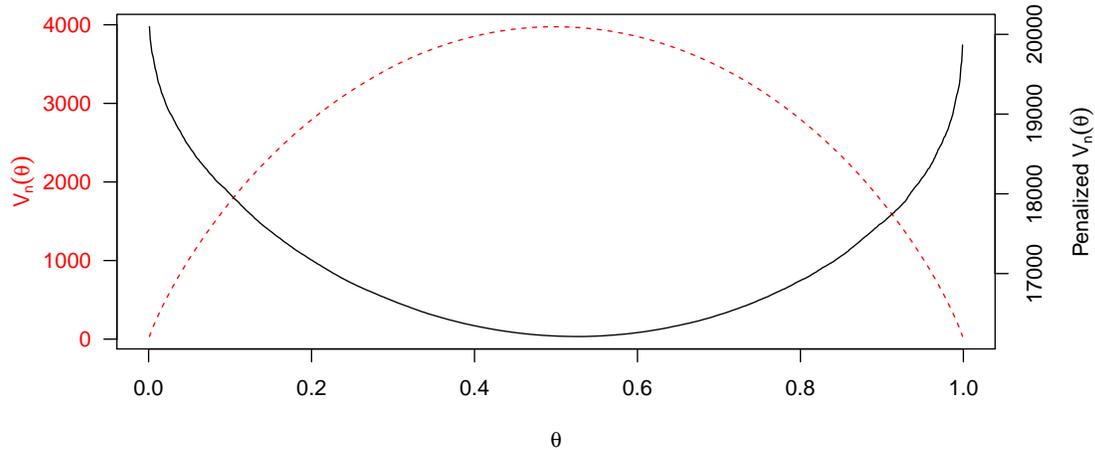}}
   \caption{\label{fig:unpenalized dispersion}Unpenalized (dashed red line) and penalized (black line) quantile dispersion for data generated from a Gaussian distribution.}
\end{figure}

In order to illustrate why this is required (focusing on $\theta$ first), consider $K=1$ and univariate data generated by a Gaussian distribution with some parameters $\mu$ and $\sigma^2$ and take $\xi=q_n(\theta)$, where $q_n(\theta)$ is the quantile computed on the sample of size $n$.
The dashed red line of Figure \ref{fig:unpenalized dispersion} shows the shape of the dispersion $D_n(\theta)=\sum_{i=1}^n \Q(x_{i},\theta,q_n(\theta))$ (w.l.o.g. $\lambda=1$) on a large sample with $n=10,000$ for a dense grid of values of the percentile between 0 and 1.

Since data have been generated by a symmetric distribution, the optimal value of $\theta$ should actually be $\frac{1}{2}$ corresponding to the median, and
Figure \ref{fig:unpenalized dispersion} shows that the penalty is required to
achieve this.

\begin{figure}[t]
  \centering
  \makebox{\includegraphics[width=\textwidth]{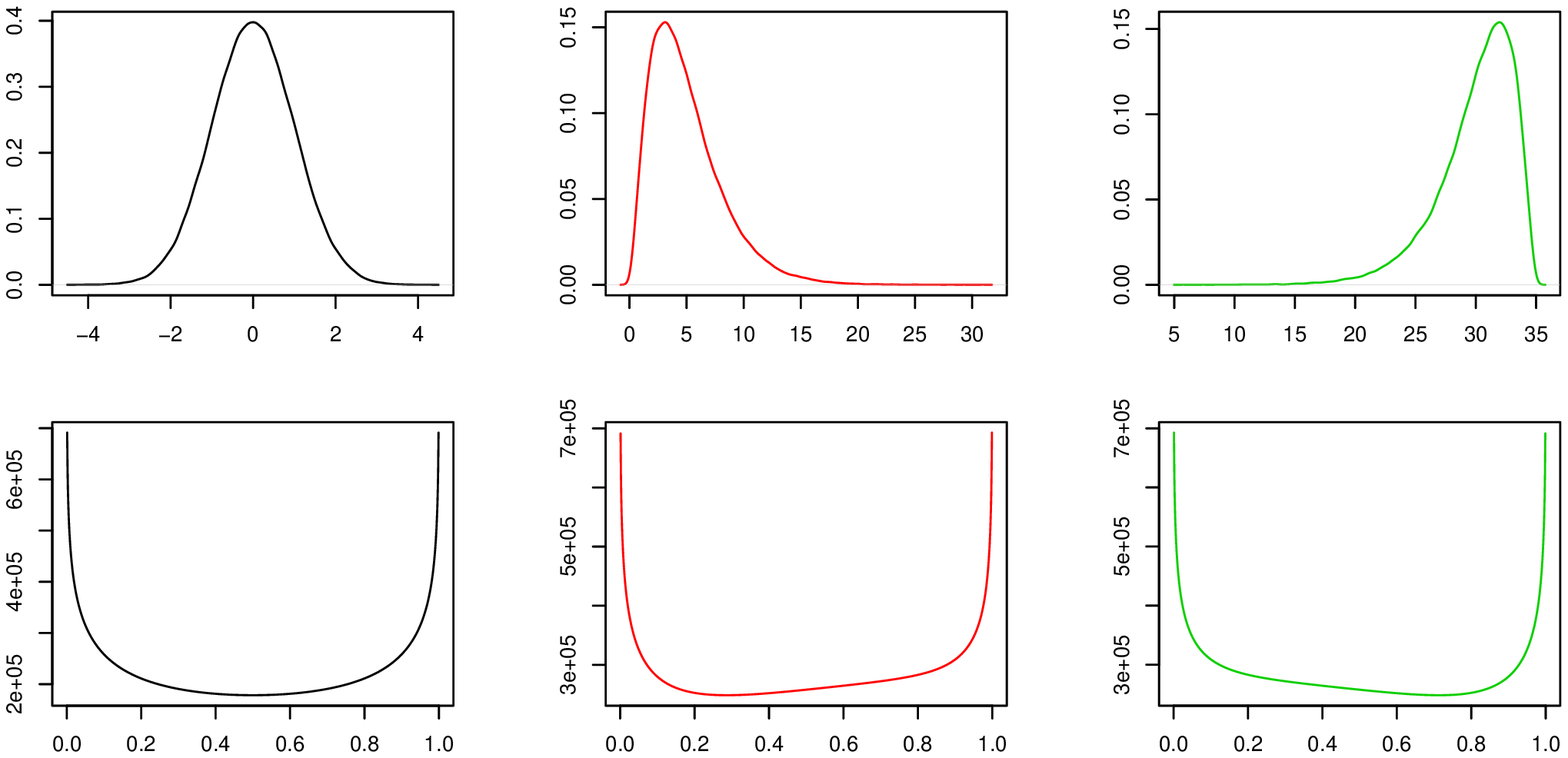}}
  \caption{\label{fig:normalizeddispersion2}In the second row of the panel the penalized dispersion function is plotted against $\theta$ for data generated according to the density functions depicted in the first row of the panel: a symmetric distribution, a positive skew distribution and a negative skew distribution.}
\end{figure}

Figure \ref{fig:normalizeddispersion2} shows the penalized dispersion function for data generated by a symmetric distribution, by a positive skew distribution and by a negative skew distribution.

The parameters $\lambda_j$ allow for implicit rescaling of the variables and scale equivariance. They need to be ``penalized'' because without the penalty term, $\blambda\to 0$ would just enforce $V_{n,K}\to 0$.

$K$-quantiles clustering
is scale equivariant, which means that the clustering remains the same, and
parameters change appropriately, if the variables in the data are multiplied
by different constants.
\begin{proposition} \label{pscale}
For constants ${\bf c}=(c_1,\ldots,c_p)^t,\ c_1,\ldots,c_p\neq 0$,
let $\tilde\x_N^*=(\x^*_1,\ldots,\x_n^*)$
be defined by $\x_i^*={\bf c}^t\x_i,\ i=1,\ldots,n$. Let
\[
T_{n,K}(\tilde\x_n)=\left(\btheta_{n,K},\tilde\bxi_{n,K},\blambda_{n,K}\right),\
{\bf d}=\left(\frac{1}{c_1},\ldots,\frac{1}{c_p}\right)^t,
\]
$\tilde\bxi^*_{n,K}=\left({\bf c}^t\bxi_{n,K,1},\ldots,{\bf c}^t\bxi_{n,K,K}\right)$.
Then,
\[
T_{n,K}(\tilde\x_n^*)=
\left(\btheta_{n,K},\tilde\bxi^*_{n,K},{\bf d}^t\blambda_{n,K}\right),
\]
and the corresponding clustering $C$ is the same as for $T_{n,K}(\tilde\x_n)$.
\end{proposition}

\medskip
The proof of Proposition \ref{pscale} can be found in Appendix \ref{proof:pscale}.

Note that the parameters $\lambda_j$ rescale the variables based on variation within clusters (only the discrepancy between $x_{ij}$ and the cluster barycenter to which $\x_i$ is assigned are taken into account, see also Proposition \ref{proplambda} below). This is more appropriate than achieving scale equivariance by standardizing the variables beforehand based on the variance or some other dispersion measure, as is sometimes done for $K$-means, see \cite{GnKeTs95}. Such methods will estimate a large dispersion if along a variable the separation between clusters is large, which may lead to downweighting of variables that are in fact very informative for clustering.

\subsection{A greedy search algorithm}\label{salgo}
Lloyd's classical $K$-means algorithm \citep{Lloyd82} is a greedy algorithm, and $K$-quantiles clustering can also be computed using a fast greedy algorithm. This is based on the following two propositions, which show that $\btheta$ and $\blambda$ minimizing $V_{n,K}$ can easily be found with all other parameters given. The propositions treat the case $p=1$ w.l.o.g., because the variables can be treated separately for minimizing $V_{n,K}$ with respect to these parameters.
\begin{proposition}\label{proptheta}
For one-dimensional $x_1,\ldots,x_n$, given $\xi_1,\ldots,\xi_K$, $C(1),\ldots,C(n)$ and $\lambda>0$, the solution to the problem
$$\theta
=\argmin_{\theta^*\in(0,1)} \sum_{i=1}^n  \lambda \Q(x_i,{\theta^*},\xi_{C(i)})-n\log (\lambda\theta^*(1-\theta^*))$$
is given by the roots of the quadratic equation
$$\theta^2 \lambda \sum_{i=1}^n(x_i -\xi_{C(i)}) -\theta \left(2n+ \lambda \sum_{i=1}^n(x_i -\xi_{C(i)})\right)+n=0.$$
\end{proposition}
\begin{proposition}\label{proplambda}
For one-dimensional $x_1,\ldots,x_n$, given $\xi_1,\ldots,\xi_K$, $C(1),\ldots,C(n)$ and $\theta\in(0,1)$, the solution to the problem
$$\lambda
=\argmin_{\lambda^*} \left(\sum_{i=1}^n \lambda^* \Q(x_i,\theta,\xi_{C(i)})-n\log (\lambda^*\theta(1-\theta)); \lambda>0 \right)$$
is given by
$$\lambda=\frac{n}{\sum_{i=1}^n  \Q(x_{i},\theta_j,\xi_{C(i)})}.$$
\end{proposition}

\medskip
Proofs of Propositions \ref{proptheta} and \ref{proplambda} are given in Appendix \ref{proof:proptheta} and \ref{proof:proplambda}, respectively.

\medskip
The greedy algorithm consists of an initialization step and a clustering step, which makes $V_{n,k}$ smaller in each step and is repeated until convergence.
Because there are only finitely many possible clusterings, the algorithm will reach convergence after a finite number of steps (as does Lloyd's algorithm).
For big datasets, if convergence takes too long, one could fix a maximum number of iterations. However, often convergence is reached very quickly; also the constrained methods proposed in Section \ref{sgen} are faster.
The scheme of the algorithm is the following:

\medskip
\noindent\rule[0.5ex]{\linewidth}{1pt}
\begin{enumerate}
  \item \emph{Initialization}: For each variable, choose randomly a value $\theta_j$ and equispaced quantiles as barycenters defined as $q_{nkj}\left(\theta^*_{kj}\right)$, with $\theta^*_{kj}=(k-1)/2(K-1)+\theta_j/2$. Set $\lambda_j=1$.
  \item \emph{Clustering step}: Repeat the following until $V_{n,K}(\boldsymbol\theta; \boldsymbol\xi)$ stops changing:
  \begin{enumerate}
    \item Compute the clustering $C(1),\ldots,C(n)$ using (\ref{eq:clustering}).
    \item For $j=1,\ldots,p$ compute $\theta_j$ using Proposition \ref{proptheta}.
    \item For $j=1,\ldots,p$ compute $\lambda_j$ using Proposition \ref{proplambda}.
    \item for $k=1,\ldots,K,\ j=1,\ldots,p$ compute the new barycenters $\xi_{kj}=q_{n_kkj}(\theta_j)$, where $n_k=\sum_{i=1}^n \indicator{C(i)=k}$, and $q_{n_kkj}(\theta_j)$ denotes the quantile among the $x_{ij}$ with $C(i)=k$.
  \end{enumerate}
\end{enumerate}
\noindent\rule[0.5ex]{\linewidth}{1pt}

\medskip
Because of (\ref{eq:clustering}) and Propositions \ref{proptheta} and \ref{proplambda}, $V_{n,K}$ is made smaller in every step, so the algorithm is guaranteed to converge. As usual, the algorithm only finds a local optimum of $V_{n,k}$. Therefore it is recommended to repeat the algorithms with a number of $h$ different initializations (default is set to 30), and the best solution is chosen according to the minimum value of $V(\boldsymbol\theta,\boldsymbol\xi)$. The algorithm is implemented in the R package \texttt{QuClu} soon available on the CRAN Web page and currently available upon request.
Note that the proposed initialization of $\blambda$ could be made scale
equivariant by for example setting $\lambda_j=1/s_j$ with $s_j^2$ being the
sample variance of variable $j$, but this may come with the same issues as
prior scaling of $K$-means, see Section \ref{spenscale}.

\subsection{Constrained versions} \label{sgen}
More parsimony and faster computation can be achieved by constraining the $\theta$ and $\lambda$-parameters. The algorithm described in Section \ref{salgo} can easily be modified to accommodate these.
\begin{itemize}
  \item Algorithm \textbf{CU}: \textbf{C}ommon $\theta$ and \textbf{U}nscaled variables. \\ A common value of $\theta$ for all the variables is assumed, and variables are not implicitly scaled (the latter can make sense in applications in which the variables have comparable meanings and measurement units, if subject matter knowledge suggests that variable importance is proportional to variation). In this case we minimize the empirical loss function:
\begin{eqnarray}\label{eqn:lossA}
V_{n,K}(\theta, \boldsymbol\xi, \tilde\x_n)=\sum_{i=1}^n \min_{k \in \{1,\ldots,K\}} \sum_{j=1}^p \Q(x_i,\theta,\xi_{kj}) -np \log (\theta(1-\theta))
\end{eqnarray}
  \item Algorithm \textbf{CS}: \textbf{C}ommon $\theta$ and \textbf{S}caled variables through $\lambda_j$. \\ A common value of $\theta$ is taken but variables are scaled through $\lambda_j$. Then the empirical loss function to be minimized is:
 \begin{eqnarray*}\label{eqn:lossB}
V_{n,K}(\theta,\boldsymbol\xi,\boldsymbol\lambda, \tilde\x_n)=\sum_{i=1}^n \min_{k \in \{1,\ldots,K\}} \sum_{j=1}^p \lambda_j \Q(x_{ij},\theta,\xi_{kj}) -n \sum_{j=1}^p\log (\lambda_j\theta(1-\theta))
\end{eqnarray*}
       \item Algorithm \textbf{VU}: \textbf{V}ariable-wise $\theta_j$ and \textbf{U}nscaled variables. \\ In this case we minimize
   \begin{eqnarray*}\label{eqn:lossC}
V_{n,K}(\boldsymbol\theta,\boldsymbol\xi, \tilde\x_n)=\sum_{i=1}^n \min_{k \in \{1,\ldots,K\}} \sum_{j=1}^p  \Q(x_{ij},\theta_j,\xi_{kj}) -n \sum_{j=1}^p\log (\theta_j(1-\theta_j))
\end{eqnarray*}
  \item Algorithm \textbf{VS}: \textbf{V}ariable-wise $\theta_j$ and \textbf{S}caled variables through $\lambda_j$. This is the most flexible method, with $V_{n,K}$ defined in (\ref{eq:defv}).
\end{itemize}
Minimisation problems CS and VS are scale equivariant (Proposition \ref{pscale}
holds with the same proof for CS as well), whereas CU and VU are not.

\section{Consistency theory}\label{sconsistency}
In this section we show that the parameters estimated by $K$-quantiles clustering from data are consistent estimators of the $K$-quantiles clustering functional, i.e., the version computed on an underlying distribution rather than on data. This essentially follows the work of \cite{Pollard81} on $K$-means clustering, although $K$-quantiles clustering has more complex parameters and therefore Pollard's original proof needs to be augmented (actually Pollard's result covers a somewhat more general clustering problem than $K$-means but this does not include $K$-quantiles clustering). It is well known in the case of $K$-means that consistency for the $K$-means functional does not imply that the estimated parameters (i.e., the $K$ mean vectors) are consistent for the parameters of the Gaussian fixed partition model for which $K$-means is the ML estimator (see \cite{BrWi78}), and in the same way the result presented here does not imply that $K$-quantiles clustering is consistent for estimating the parameters of a fixed partition model of asymmetric Laplace distributions as introduced in Section \ref{sfp}. Anyway, the consistency result given here is essentially nonparametric, for very general distributions, and it ensures the asymptotic stability of $K$-quantiles clustering, and the estimated parameters can be analyzed by considering the $K$-quantiles clustering functional. There is some literature on convergence rates and ``performance guarantees'' for $K$-means clustering (e.g., \cite{LiLuZe94,ORSS06}), but this relies on strong assumptions, and generalizing such results to $K$-quantiles clustering is beyond the scope of the present work. Instead, after the consistency result in Theorem \ref{tcons}, we show in Theorem \ref{tlevel} that the $K$-quantiles clustering functional defines clusters that are in line with ``central sets'' in a nonparametric mixture situation with strong separation between mixture components.

We consider the most flexible and general model defined above, with variable-wise $\theta$ and scaled variables, which is the most difficult one for proving consistency. Corresponding results for the less flexible models can be obtained more easily.

The proof relies heavily on showing that parameter estimators for large $n$ do not leave a compact set, but (considering a single variable) $\lambda\to\infty$ and $\theta\to 0$ or $\theta\to 1$ may happen together without constraints on the parameter space (leading to the exponential distribution in the limit, which in practice could actually be integrated in the approach), causing trouble with uniform convergence arguments. This can be avoided by either constraining $\theta_j\in[r,1-r],\ r>0$, or $\lambda_j\le \lambda^+<\infty$ for $j\in\{1,\ldots,p\}$. We will impose the latter constraint here, so that results hold without constraint for the unscaled case, i.e., $\blambda={\bf 1}$.

The parameter space used here is
\[
\S=\{(\btheta,\tilde\bxi,\blambda):\ \theta_j\in (0,1),
\bxi_k\in \R^p, \lambda_j\in (0,\lambda^+],\ j\in\{1,\ldots,p\},\ k\in\{1,\ldots,K\}\}.
\]
We use the notation defined in (\ref{eq:defv}) and (\ref{eq:estimation});
in case that the argmin is not unique, any solution can be taken.
We modify (\ref{eq:defv}) multiplying by $\frac{1}{n}$ in order
to use stochastic convergence
of means to expectations:
\[
V_{n,K}(\Theta,\tilde\x_n) = \frac{1}{n}\sum_{i=1}^n
\sum_{j=1}^p \lambda_j \Q(x_{ij},\theta_j,\xi_{C(i)j}) -\sum_{j=1}^p \log \lambda_j\theta_j(1-\theta_j).
\]
For a given distribution $P$ on $\R^p$ define
\begin{eqnarray*}
V_{K}(\btheta,\tilde\bxi,\blambda,P)&=&\int \min_{k \in \{1,\ldots,K\}} \sum_{j=1}^p \lambda_j \Q(x_{j},\theta_j,\xi_{kj})dP(\x) -\sum_{j=1}^p\log \lambda_j\theta_j(1-\theta_j),\\
T_K(P)&=&(\btheta_{K},\tilde\bxi_{K},\blambda_{K})=\argmin_{(\btheta,\tilde\bxi,\blambda)\in\S}V_{K}(\btheta,\tilde\bxi,\blambda,P).
\end{eqnarray*}
Let $S_{n,K}=V_{n,K}(T_{n,K}(\tilde\x_n),\tilde \x_n),\ S_K=V_K(T_K(P),P)$.
In order to avoid issues due to label switching of the clusters, we consider
consistency of lists $(\btheta,\tilde\bxi,\blambda)$, where
$\tilde\bxi$ is the set of quantiles.
Convergence and continuity are
defined in terms of a distance $d$
between two such lists $(\btheta_1,\tilde\bxi_1,
\blambda_1),  (\btheta_2,\tilde\bxi_2,\blambda_2)$
that is the maximum of $\|\btheta_1-\btheta_2\|, \|\blambda_1-\blambda_2\|$,
and the maximum over the Euclidean distances between any element of
$\tilde\bxi_i$ and its closest element of $\tilde\bxi_j$, $i\neq j, i,j=1,2$
(known as Hausdorff distance between $\tilde\bxi_1$ and $\tilde\bxi_2$).

The following assumptions will be required:
\begin{description}
\item[A1] $B=\int \|\x\| dP(\x)<\infty$.
\item[A2] $T_k(P)$ is uniquely defined (up to cluster labelling) for
$k=1,\ldots,K$.
\end{description}

A1 means that all involved integrals are finite; note that \cite{Pollard81}
requires $\int \|\x\|^2 dP(\x)<\infty$ for $K$-means. A2 enforces stability;
as \cite{Pollard81} noted for $K$-means, it implies that
$S_K<S_{K-1}<\ldots<S_1$ because if $S_k=S_{k-1}$ for some $k$, one could
add any point to $\tilde\bxi_{k-1}$ to construct $\tilde\bxi_{k}$ that cannot
have a worse value than $S_k$ together with $\btheta_{k-1}, \blambda_{k-1}$.

\begin{theorem} \label{tcons}
If $\x_1,\x_2,\ldots \sim P$ i.i.d., and assumptions A1 and A2 hold,
then, for $n\to\infty$:  $T_{n,K}(\tilde\x_n)\to T_K(P),\ S_{n,K}\to S_K$ a.s.
\end{theorem}

\medskip
The proof of Theorem \ref{tcons} is given in Appendix \ref{proof:tcons}.

\medskip
The value of $T_K(P)$ for given $P$ implies a clustering of $\R^p$ by
\[
\gamma_{T_K(P)}(\x)=\argmin_k \sum_{j=1}^p \lambda_{Kj} \Q(x_{j},\theta_{Kj},\xi_{Kkj})
\]
for $\x=(x_1,\ldots,x_p)\in\R^p$. The next result is about this implied
clustering in case that $P_m,\ m\in\N$ is a sequence of mixture distributions
with mixture components between which
the separation becomes larger and larger
with increasing $m$.

Here are some definitions and assumptions.
Let $G_1,\ldots,G_K$ be distribution functions on $\R^p$ defining distributions
$Q_1,\ldots,Q_K$
parameterized in such a way that 0 is their ``center'' in some sense; it
could be the mode, the mean, the multivariate median or quantile; important
is only that $G_i$ is defined relative to 0.
Let $\pi_1,\ldots,\pi_K>0$ mixture proportions with $\sum_{k=1}^K\pi_k=1$.
Assume
\begin{description}
\item[A3] For $m\in\N,\ k\in\{1,\ldots,K\}$ let
$\brho_{mk}\in \R^p$ sequences so that
\[
\lim_{m\to\infty}\min_{k_1\neq k_2\in\{1,\ldots,K\}}\|\brho_{mk_1}-\brho_{mk_2}\|
=\infty.
\]
\item[A4] $\exists B_0<\infty$ so that for all $k\in\{1,\ldots,K\}\
:\ \int\|\x\|dG_k(\x)\le B_0.$
\end{description}
Assumption A3 enforces the distance between central sets
to become larger and larger.
A4 makes sure the involved expectations exist (note that a similar theorem could
be proved for $K$-means, but this would require a bound on the mixture
component-wise $E\|\x\|^2$).

Define a sequence of distributions $P_m$ with distribution functions
$F_m$ on $\R^p$ by
$F_m(\x)=\sum_{k=1}^K\pi_kG_k(\x-\brho_{mk})$. Consider, for $\epsilon>0$,
the ``central set'' $\{\x:\ \|\x\|<\epsilon\}$ about 0.  Then, by choosing
$0<\epsilon<\infty$ large enough,
\begin{equation}\label{eq:a5}
\exists \delta>0:\
\forall k\in\{1,\ldots,K\}:\
\pi_k Q_K\{\|\x\|<\epsilon\} \ge \delta.
\end{equation}
The following theorem states that in this setup, when evaluating the
$K$-quantiles clustering functional, eventually the different clusters include
the full central sets of the different mixture components (and central sets
can be of arbitrarily large though fixed radius), and in this sense
the clustering corresponds to the mixture structure.
The mixture components are allowed to overlap, although
for $m\to\infty$ the overlap becomes arbitrarily small.
\begin{theorem}\label{tlevel} With the above definitions, assuming A3 and A4,
for large enough $m$, the clusters of $T_K(P_m)=(\btheta_{m},\tilde\bxi_{m},\blambda_{m})$ can be numbered in such a
way that for $k\in\{1,\ldots,K\}:$
\[
\{\x:\ \|\x-\brho_{mk}\|<\epsilon\} \subseteq \{\x:\ \gamma_{T_K(P_m)}(\x)=k\}.
\]
\end{theorem}

The proof of Theorem \ref{tlevel} is given in Appendix \ref{proof:tlevel}.

 \section{Simulation Study} \label{sec:sims}

The performance of the $K$-quantiles clustering algorithm is evaluated in an extensive simulation study. We generate $p$ vectors from $K$, $K=2,3,5$, populations, $\mathbf{X}^{(K)}$, according to five different scenarios:

\begin{enumerate}
\item In the first scenario, we consider symmetric Student $t$-distributed variables $W_j$ ($ j =1, \ldots, p$) with three degrees of freedom, and we simulate $K$ location-shifted populations from $W_j$, each shift from the closest population being unitary [e.g. $X_j^{(1)} \sim W_j$, $X_j^{(2)} \sim (W_j+1)$, $X_j^{(3)} \sim (W_j-1)$, \ldots].

\item In the second scenario,
we test the behaviour of the clustering algorithm in highly skewed data by generating identically distributed
vectors $W_j$ ($j =1, \ldots, p$) from a multivariate zero-centered Gaussian distribution, transforming them
using the exponential function and shifting contiguous populations by 0.6 [e.g. $X_j^{(1)} \sim exp(W_j)$, $X_j^{(2)} \sim (exp(W_j)+0.6)$, $X_j^{(3)} \sim (exp(W_j)-0.6)$, \ldots].

\item In the third scenario, we
consider different distributions for the $p$ variables. We first generate $W_j$ from a multivariate
zero-centered Gaussian distribution and then split $p$ into five balanced blocks to which we apply different transformations:
\begin{itemize}
\item[(i)] a location shift [e.g. $X_j^{(1)} \sim W_j$, $X_j^{(2)} \sim (W_j+0.7)$, $X_j^{(3)} \sim (W_j+1.4)$, \ldots];
\item[(ii)] an exponential transformation on the shifted populations at (i) [e.g. $X_j^{(1)} \sim exp(W_j)$, $X_j^{(2)} \sim exp(W_j+0.7)$, $X_j^{(3)} \sim exp(W_j+1.4)$, \ldots];
\item[(iii)] a logarithmic transformation on the shifted populations at (i) [e.g $X_j^{(1)} \sim log(|W_j|)$,  $X_j^{(2)}  \sim  log(|W_j +0.7|)$, $X_j^{(3)}  \sim  log(|W_j +1.4|)$,\ldots];
\item[(iv)] a quadratic transformation on the shifted populations at (i) [e.g $X_j^{(1)} \sim W_j^2$, $X_j^{(2)} \sim (W_j+0.7)^2$, $X_j^{(3)} \sim (W_j+1.4)^2$, \ldots];
\item[(v)] a square root transformation on the shifted populations at (i) [e.g. $X_j^{(1)} \sim \sqrt{|W_j|}$, $X_j^{(2)} \sim \sqrt{|W_j+0.7|}$, $\X_j^{(3)} \sim \sqrt{|W_j+1.4|}$, \dots].
\end{itemize}

\item In the fourth scenario, we simulate different distributional shapes and
levels of skewness even for different classes within each variable. Within each class, data
are generated according to beta distributions, $X_j^{(k)} \sim Beta(a,b)$, $j=1,\ldots,n$ and $k=1,\ldots,K$, with parameters $a$ and $b$ in the interval $(1,10)$
randomly generated for each class within each variable. The absolute difference between the class expected values for each variable is bounded from above by 0.2 (this is done in order to not make the clustering task too easy; as cluster differences are aggregated over many dimensions, simulated clusters may be so strongly separated that every method can find them easily).

\item The fifth scenario is similar to the fourth one.
Within each class, data
are generated according to beta distributions, $X_j^{(k)} \sim Beta(a,b)$, $j=1,\ldots,n$ and $k=1,\ldots,K$, with parameters $a$ and $b$ randomly chosen to be in the intervals: $(0,1)$ and $(1,5)$, or $(0,1)$ and $(1,10)$, $(1,3)$ and $(5,10)$,$(1,3)$ and $(1,3)$ so as to guarantee a higher level of skewness for some variables, for each class within each variable. The absolute difference between the class expected values for each variable is bounded from above by 0.1.
\end{enumerate}

For each of the five scenarios and for each set of $K$ populations, $K=2,3,5$, we evaluate combinations of $p=50$, $100$, $500$, $n=50$,
$100$, $500$, different percentages of relevant variables for grouping structure, i.e., $100\%$, $50\%$ and $10\%$, independent or dependent variables (scenario (a)-(c) only), for a total of 648 different settings. In the ``dependent variables'' case, a dependence structure
is introduced by generating variables $W_1,\ldots,W_p$
from either a $t$ or a Gaussian distribution with random correlation matrix based on the method proposed by
\cite{Joe06}, so that the correlation matrices are uniformly distributed over the space of positive definite
correlation matrices, with each correlation marginally distributed as $Beta(p/2, p/2)$ on the interval
$(-1, 1)$. The irrelevant noise variables are generated independently of each other from the base distribution of each scenario. For each setting we simulate 100 datasets and we record the Adjusted Rand Index (ARI) \citep{Hubert85} of the yielded classification compared with the true cluster membership.

We compare the $K$-quantile clustering algorithms' capability of recovering the original cluster memberships with those of seven other clustering methods: two model-based clustering approaches (mixture of Gaussians, mixture of factor analyzers) \citep{Peel}, $K$-means algorithm \citep{Lloyd82}, Partition Around Medoids \citep{Kaufman05}, agglomerative hierarchical clustering with unweighted pair group method (UPGMA), spectral clustering \citep{ng2002spectral} and affinity propagation \citep{Frey972}. The inclusion of irrelevant variables may prompt the idea that also clustering methods with variable selection should be tried out; however, variable selection is usually defined on top of an existing clustering method without variable selection, see, e.g., \cite{FrMe04}. Such ideas can be applied to $K$-quantiles clustering as well as to the competing methods, which we leave for future research. Mixture of Skew-$t$ distributions were also considered; however, as solutions were available only 20\% of times, such results are here discarded.

Details about the implementation and parameter tuning of these methods are given in the Supplementary Material, where also detailed tables on simulation results are reported.

\begin{figure}
\centering
\includegraphics[width=0.9\textwidth]{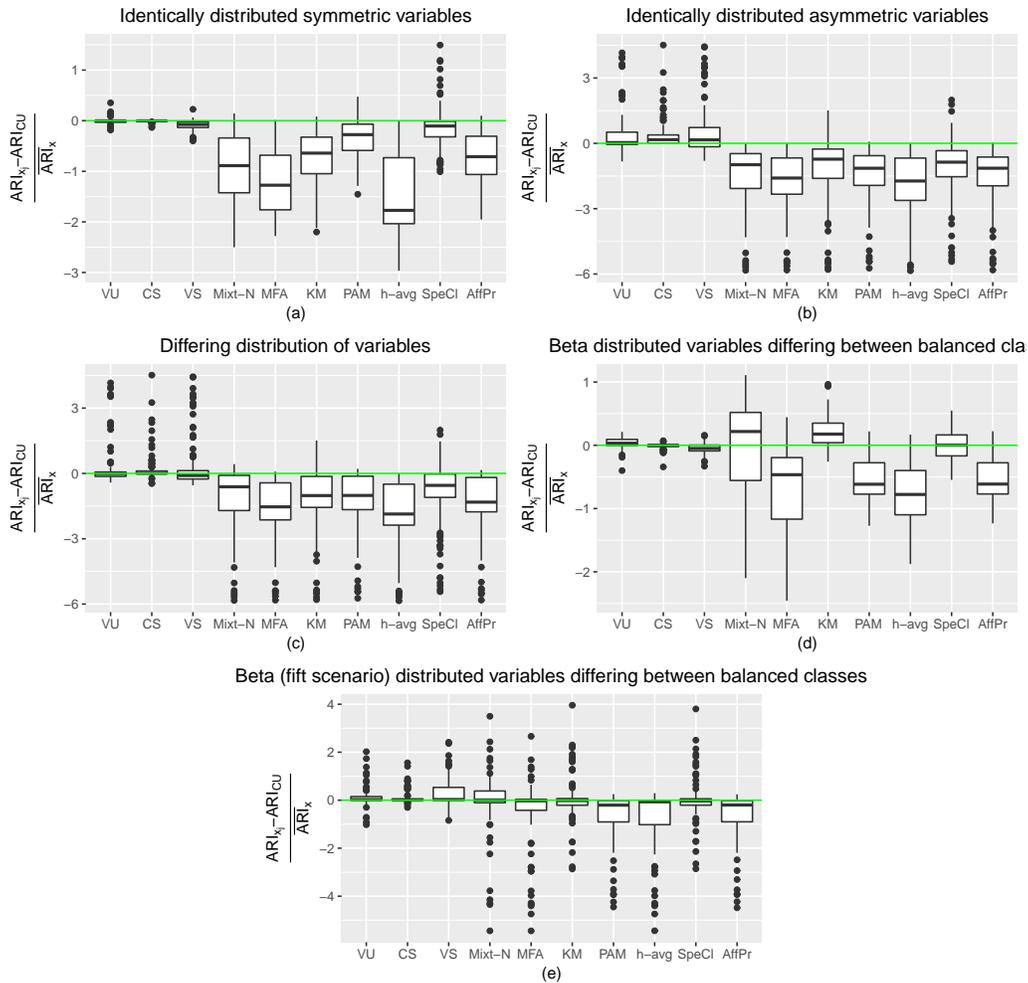}
\caption{\label{fig:boxplot}Performance of different clustering algorithms relatively to the Common Unscaled $k$-quantile clustering algorithm. The labels along the horizontal axis refer to the different methods: CS, Common Scaled $k$-quantile; VU, Variable-wise Unscaled $k$-quantile; VS, Variable-wise Scaled $K$-quantile; \emph{Mixt-N}, mixture of Normal distributions; \emph{MFA}, mixture of factor analyzers; \emph{KM}, $k$-means clustering; \emph{PAM}, Partition Around Medoids algorithm; \emph{h-avg}, hierarchical clustering with average linkage (UPGMA); \emph{SpeCl}, spectral clustering; \emph{AffPr}, affinity propagation. The five panels show the distribution of the Adjusted Rand Index for (a) identically distributed symmetric variables, (b) asymmetric variables, (c) different distributions of variables, (d) different distributions of classes within variables in balanced and unbalanced populations and (e) different (skew) distributions of classes within variables in balanced and unbalanced populations.}
\end{figure}

More specifically, we evaluate the accuracy of each clustering method as its ARI minus the ARI of the
Common Unscaled $K$-quantile clustering algorithm, divided by the average ARI in the given setting (for computing this average, the three percentages of relevant variables are aggregated in order to avoid blowing up small differences between uniformly small ARI values for only 10\% relevant variables by small denominators; where methods do not deliver a solution, the ARI has been set to zero). This is done for the sake of enabling a simpler display of the many results, because it allows to aggregate results for different $K$, $n$, $p$, dependence structure, and percentage of relevant variables by scaling all these results so that they become comparable. Raw ARI values are given in the Supplementary Material. The aggregated
distributions of these rescaled results are displayed in the boxplots of Figure \ref{fig:boxplot}.

For all methods, the capability of recovering the original cluster membership improves as the sample size increases. For the $K$-quantile clustering, this is particularly evident in the scenarios where the percentage of relevant variables is very low; in all the other cases, in fact, results are generally very good and no remarkable difference due a different sample size can be noticed.
All the methods, for
fixed sample size and percentage of relevant variables, seem to perform better as
$p$ increases in almost all of the settings. As could be expected, clustering performances worsen as the number of irrelevant variables increases.

The $K$-quantile methods perform very well in most situations compared to other clustering approaches.
In the scenarios with identical distributional shapes and symmetric variables, solutions from the quantile clustering are mostly preferable to those from any other method. Not surprisingly, common $\theta$ quantile procedures, i.e. CU and CS, slightly outperform those with variable-wise $\theta_j$s.

In the settings with identical distributional shapes and asymmetric variables, $K$-quantile clustering methods outperform
all other methods clearly and more or less uniformly; here, procedures with a variable-wise $\theta_j$, i.e. VU and VS, seem to produce a slightly better clustering.

With different distributions of variables, the $K$-quantile clustering methods again show very good global results. Only occasionally, in some situations with just 10\% relevant variables, Gaussian mixtures, $K$-means and spectral clustering can improve on $K$-quantiles.

In the fourth scenario with beta distributions differing between variables and classes within variables, the $K$-quantile clusterings do not always outperform the other methods: while they generally produce good results, they often fall behind the accuracy of the $K$-means algorithm, spectral clustering, and also Gaussian mixtures. The reason why this happens is that although these distributions are skew, their tails vanish outside the unit interval, and often the difference between means is the most distinct feature discriminating the clusters. Therefore a squared loss function is suitable for finding them. This contrasts with the fact that $K$-quantiles beats these methods for symmetric but $t$-distributed data in the first scenario, despite the fact that $K$-means and Gaussian mixtures implicitly assume symmetry, as opposed to $K$-quantiles; however, the squared loss function is more affected by outliers in these cases.

The results of the fourth scenario prompted us to set up the fifth one, with parameters of the beta random variables chosen from different intervals so that there is more extreme skewness, and information about clustering is rather connected to distributional features other than the means. In this situation, the $K$-quantile VS algorithm is the best. $K$-means and spectral clustering still yield fairly good results, although worse than the $K$-quantiles algorithms. Gaussian mixtures also still do well, probably because flexible covariance matrices are still versatile here to adapt to these setups. Their median performance is about on a par with three of the four $K$-quantiles algorithms (results vary depending on whether $p$ is rather large compared to $n$ or not, see the Supplementary Material) but worse than the VS algorithm.

Generally, the capability of recovering the clustering memberships and the rankings of the methods do not change much with dependence, although performances are slightly better under independence. Similarly, the ranking of the methods does not strongly depend on the number of clusters.

The Supplementary Material provides some information on computing times.
Currently our implementation for running the $K$-quantiles is coded in R;
faster implementations are certainly possible. However, our experiments show
that the growth in computation time with $n$ is much slower than for PAM and
the mixture model-based methods, so that for the largest data format that we
tried ($n=50000, p=100$) our $K$-quantiles implementation is substantially
faster than all mixtures and PAM, beaten only by $K$-means (the hierarchical clustering does not deliver a solution). This demonstrates
that $K$-quantiles have the potential to be used with very large datasets.

 \section{Application to Gene Expression Data}\label{sgene}
For illustration, we apply the $K$-quantile clustering algorithms to gene expression data from the leukaemia microarray study of \cite{Golub99}.
The dataset contains the expression levels of 3051 genes for 38 leukaemia patients, obtained from acute leukaemia patients at the
time of diagnosis. The study reports that 27 subjects have Acute Lymphoblastic Leukaemia (ALL), while 11 have Acute Myeloid Leukaemia (AML). The objective is to group the set of 38 patients so as to reflect the corresponding leukaemia diagnosis by employing information coming from their gene expression levels.

Data are taken from the R package \texttt{plsgenomics} and are analysed by the same clustering methods described in Section \ref{sec:sims}. The number of true clusters for all the methods is taken as known and set equal to 2. As different versions of Mclust delivered different results, we have tried out different initializations and we chose the one with largest likelihood. For $K$-means five random starts are run. The default settings of all the other algorithms are considered.

\begin{figure}
\centering
\includegraphics[width=\textwidth]{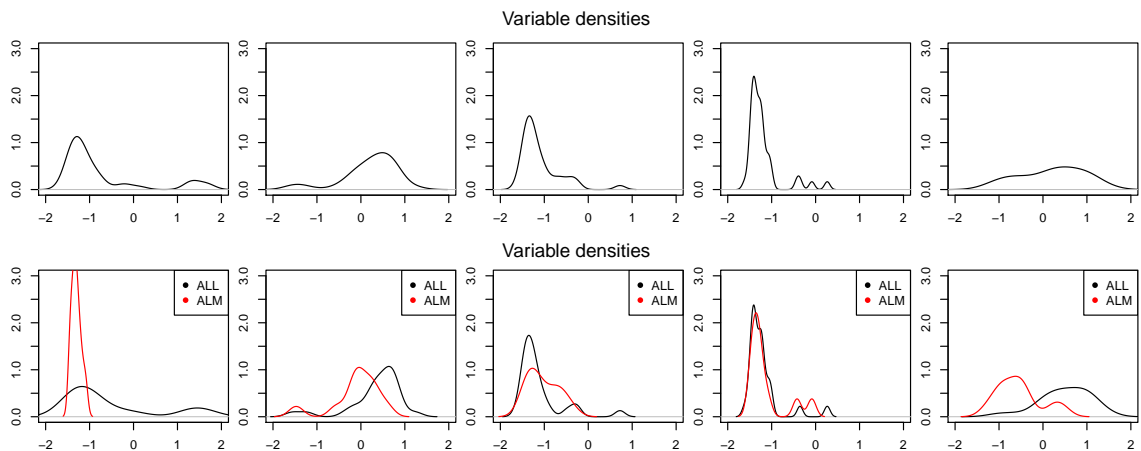}
\caption{\label{fig:density_leuk}Leukaemia dataset: densities of five randomly
selected variables (gene expression levels; fitted by R's density function with
default settings). First row: all observations together. Second row: by true cluster.}
\end{figure}

Results from all the other methods are shown in Table \ref{tab:Leukemia}. As can be seen, the $K$-quantile clustering algorithm with variable-wise $\theta_j$ and scaled variables via $\lambda_j$ is able to perfectly recover the original clustering memberships. When using the unscaled version, the performance of VU is still pretty good and superior to the other solutions.
Quantile methods with common $\theta$, whether using the scaled or the unscaled version, are not able to detect the grouping structure identified by the diagnosis: this is probably due to the fact that the distribution of the expression levels is really different for different genes (see Figure \ref{fig:density_leuk}, where the density of a random sample of gene expression levels is plotted).

Mixture of Factor Analyzers could not return any solution, as it has ended up with errors.

Mixture of Gaussians and $K$-means algorithm yield exactly the same clustering (up to label switching); their results are still good. As the number of variables is very large, \texttt{Mclust} could only estimate mixture of Gaussians with spherical or diagonal covariance matrices, reducing to 6 out of 14 possible parsimonious models, namely: \texttt{EII} (spherical, equal volume), \texttt{VII} (spherical, unequal volume), \texttt{EEI} (diagonal, equal volume and shape), \texttt{VEI} (diagonal, varying volume, equal shape), \texttt{EVI} (diagonal, equal volume, varying shape), \texttt{VVI} (diagonal, varying volume and shape).

Spectral clustering and affinity propagation provide overall good results but worse than the mixture of Gaussians.

\begin{table}
  \caption{\label{tab:Leukemia} Adjusted Rand Index multiplied by 100 for the leukaemia data set.}
  \centering
\begin{tabular}{lr}
\hline
 Method & ARI \\
 \hline
  CU & 3.28 \\
  VU & 100.00 \\
  CS & -2.61 \\
  VS & 89.13 \\
  Mixture of Normals & 79.27 \\
  Mixture of FA & NA  \\
  $k$-means & 79.27 \\
  pam & 61.20 \\
  h-avg & -3.06 \\
  SpeCl & 69.92  \\
  AffPr & 61.20  \\
\hline
 \end{tabular}
\end{table}

\section{Conclusion}\label{sconc}
$K$-quantiles clustering is a new clustering method based on representing clusters by quantiles of the within-cluster distribution. It can be interpreted as Maximum Likelihood estimator for a fixed partition model of asymmetric Laplace distributions, but like $K$-means it is not in the first place meant to be associated with a specific model assumption, but rather to provide an intuitive objective function that allows for within-cluster skewness and is easy to optimize locally using a Lloyd-type algorithm. In our simulations the method did well on a wide range of within-cluster models different from the asymmetric Laplace.

\cite{Hennig15} encourages researchers to give potentially informal descriptions of what specific kinds of clusters a new clustering method is meant to find. The development of $K$-quantiles clustering was motivated in the first place by the potential of the quantile-based discrepancy to add flexibility to $K$-means, particularly regarding within-cluster skewness. The underlying model suggests that clusters can be distributions of which the marginals are unimodal and potentially skew; Theorem \ref{tlevel} shows that sufficiently well separated subpopulations will be $K$-quantile clusters even if not unimodal (as long as $K$ is fixed and there are more than $K$ modes, it is not possible to have only unimodal clusters). Similarly to $K$-means, the $K$-quantiles objective function sums up information over the different variables. This does not necessarily mean that variables have to be independent within clusters, but information about dependence is not used. The discrepancy is just an aggregation of variable-wise information. Clusters will not be rotation invariant and information carried in the original variables will be lost when considering linear combinations such as principal components. The clusters are treated as of the same shape, although with enough separation this does not stop the method from finding clusters with different shapes, see Theorem \ref{tlevel} and the simulations.
The advantage of this is that a parsimonious parametrization allows the handling of high-dimensional data. An obvious generalization would be to allow the parameters $\theta$ and $\lambda$ to vary between clusters, but this is likely to require considerably more computational effort.
The use of unsquared distances gives outliers less influence on the cluster barycenters than in $K$-means or ML-estimators for Gaussian mixtures.

The number of clusters $K$ is fixed here, and estimating $K$ is left to future work. Many methods for estimating the number of clusters are based on computing a clustering for a range of values of $K$ and then cluster validation indexes or stability assessments are used to pick the best $K$ \citep{HVH16,Leisch16}. Such an approach can be used for estimating $K$ together with $K$-quantiles clustering in the same manner as with $K$-means or $K$-medoids. Similarly, principles for variable selection that exist for $K$-means and other clustering methods could be applied to $K$-quantiles.

The penalization of the quantile defining probability $\theta$ and the scaling parameter $\lambda$ as derived here from a fixed partition model of asymmetric Laplace distributions may also be helpful for quantile-based supervised classification as introduced in \cite{HeVi16}.


\bibliographystyle{chicago}

\bibliography{kquantilesref}

\newpage

\appendix
\footnotesize
\section{Appendix: Proofs of propositions and theorems}
\subsection{Proof of Proposition \ref{pquantile}} \label{proof:pquantile}
The sum in (\ref{eqn:prop1}) can be minimized for each component independently,
see (\ref{eqn:qdistp.centroid}), and also separately for $k=1,\ldots,K$.
Therefore consider w.l.o.g. $p=1$ and $K=1$.

Then the right side of (\ref{eqn:prop1}) can be written as
$$ \sum_{x_i \leq \xi } (1-\theta) (\xi - x_i) +  \sum_{x_i > \xi } \theta (x_i -\xi ) $$
and the score function is
\[
\sum_{x_i \leq \xi } (1-\theta) -  \sum_{x_i > \xi } \theta= \sum_{i=1}^n \indicator{x_i \leq \xi}-n\theta,
\]
which is zero for $\theta=\frac{1}{n}\sum_{i=1}^n \indicator{x_i \leq \xi}$, so that $\xi=q_n(\theta)$ (any of the possible interval of quantiles).

\subsection{Proof of Proposition \ref{pscale}}\label{proof:pscale}
For all $\Theta=(\btheta,\tilde\bxi,\blambda,C)$ from the parameter
space:
\[
V_{n,K}(\Theta,\tilde\x_n)=V_{n,K}(\btheta,\tilde\bxi^*,{\bf d}^t\blambda,C,
\tilde\x_n^*)+n\sum_{j=1}^p\log c_j,
\]
where $\tilde\bxi^*=\left({\bf c}^t\bxi_{1},\ldots,{\bf c}^t\bxi_{K}\right)$.
As $n\sum_{j=1}^p\log c_j$ is a constant for a given dataset,
minimizers of
$V_{n,K}(\btheta,\tilde\bxi,\blambda,C,
\tilde\x_n^*)$ are obtained from minimizers of
$V_{n,K}(\btheta,\tilde\bxi,\blambda,C,\tilde\x_n)$ in the required way.

\subsection{Proof of Proposition \ref{proptheta}}\label{proof:proptheta}
The proof derives by taking the first derivative of $\sum_{i=1}^n  \lambda \Q(x_i,{\theta^*},\xi_{C(i)})-n\log (\lambda\theta^*(1-\theta^*))$ with respect to $\theta^*$, which gives:
\begin{eqnarray*}
  &&\frac{\partial}{\partial\theta^*}\left(\lambda \sum_{i=1}^n  \left\{\theta^*+(1-2\theta^*)\indicator{x_i<\xi_{C(i)}} \right \}|x_i-\xi_{C(i)}| - n \log (\lambda \theta^* (1-\theta^*)) \right) = \\
  &&\lambda \sum_{i=1}^n(x_i -\xi_{C(i)}) - \frac{(1-2\theta^*)n}{\theta^*(1-\theta^*)}.
\end{eqnarray*} By equating the previous expression to zero and by multiplying by $-\theta^*(1-\theta^*)$ we get the quadratic solution for $\theta$.

\subsection{Proof of Proposition \ref{proplambda}}\label{proof:proplambda}
Similarly to proposition 2, the proof derives by computing the score with respect to $\lambda^*$:
\begin{eqnarray*}
  &&\frac{\partial}{\partial\lambda^*}\left(\lambda^* \sum_{i=1}^n  \Q(x_i,\theta,\xi_{C(i)}) - n \log (\lambda^* \theta (1-\theta)) \right) = \\
  && \sum_{i=1}^n  \Q(x_i,\theta,\xi_{C(i)}) - \frac{n}{\lambda^*}=0.
\end{eqnarray*}

\subsection{Proof of Theorem \ref{tcons}}\label{proof:tcons}
The principle of the proof is to show that
$T_{n,K}(\tilde\x_n)$ for large enough $n$ has to lie in a compact set
${\cal C}$.
In this compact set, by the uniform law of large numbers,
$V_{n,K}(\btheta,\tilde\bxi,\blambda,\tilde\x_n)$ will converge uniformly to
$V_{K}(\btheta,\tilde\bxi,\blambda,P)$, which in turn, together with continuity,
will also enforce
the minimizer to converge. $(\btheta,\blambda)$ optimizing $V_{n,K}$
are enforced to eventually lie in a compact set by the penalty term
$-\log \lambda\theta(1-\theta)$. For $\tilde\bxi$, the argument is
inductive, similar to what was done
in \cite{Pollard81}.
It is first shown that at least one of the optimizing
$\bxi_k$ must lie in a compact set, and then, assuming that this holds for
$K-1$ clusters but not for $K$, the $K$th cluster can be shown to have an
asymptotically negligible
additional contribution to $S_K$ so that $S_K=S_{K-1}$ with contradiction
against A2.

In order to show that of $T_{n,K}(\tilde\x_n)=(\btheta_n,\tilde\bxi_n,\blambda_n)$ eventually $(\btheta_n,\blambda_n)$ and at least one of the $\bxi_k$ must lie
in a compact set, define $(\btheta_0,\tilde\bxi_0,\blambda_0)$ as follows.
For $j=1,\ldots,p,\ k=1,\ldots,K$,
$\theta_{0j}=\frac{1}{2},\ \lambda_{0j}=1,\ \xi_{0kj}=0$.
Then,
\[
V_{n,K}(\btheta_0,\tilde\bxi_0,\blambda_0,\tilde\x_n)=
\frac{1}{n}\sum_{i=1}^n \sum_{j=1}^p \frac{1}{2}|x_{ij}| -p\log \frac{1}{4}.
\]
The first part converges a.s. to $\frac{B_1}{2}$, where
$B_1=\int\sum_{j=1}^p|x_j|dP(\x)\le \sqrt{p}B<\infty$ as defined in A1.

Suppose that (at least for a subsequence; apply
this qualification also to further
limits below, where necessary)
$\lambda_{nj}\to 0$, $\theta_{nj}\to 0$ or $\theta_{nj}\to 1$.
In this case
$-\log \lambda_{nj}\theta_{nj}(1-\theta_{nj})\to \infty$, and eventually
\[
V_{n,K}(\btheta_n,\tilde\bxi_n,\blambda_n,\tilde\x_n)> \frac{\sqrt{p}B}{2}- p\log \frac{1}{4},
\]
for which reason
$\lambda_{nj}\to 0$, $\theta_{nj}\to 0$ or $\theta_{nj}\to 1$
cannot happen when minimizing $V_{n,K}$. Therefore
$\exists \theta^->0, \lambda^->0$ so that for large enough $n$, a.s.,
$\min(\theta_{n1},\ldots,\theta_{np},1-\theta_{n1},\ldots,1-\theta_{np})\ge \theta^-,\
\min(\lambda_{n1},\ldots,\lambda_{np})\ge \lambda^-$.

Consider a compact set $M\subset \R^p$ with $0\in M,\ P(M)>0,\
|x_j|\le m<\infty$ for $\x\in M$.
Now suppose that there is no compact interval $\Xi$ so that
for large enough $n$, with suitable numbering of the clusters,
at least for one $k\in\{1,\ldots,K\}:\ \xi_{nk1},\ldots \xi_{nkp}\in \Xi$.
Therefore, $\xi^-_n=\min_{k\in\{1,\ldots,K\}}\max_{j\in\{1,\ldots,p\}}|\xi_{nkj}|\to\infty$
and, for $\x\in M:$
\[
\min_{k\in\{1,\ldots,K\}}\sum_{j=1}^p \lambda_j \Q(x_j,\theta_j,\xi_{kj})
\ge \sum_{j=1}^p \lambda^- \theta^-(1-\theta^-)(\xi^-_n-m).
\]
For large enough $n$ this would make
\[
V_{n,K}(\btheta_n,\tilde\bxi_n,\blambda_n,\tilde\x_n)\ge P(M)p\lambda^- \theta^-(1-\theta^-)(\xi^-_n-m)-p\log \frac{1}{4} >
\frac{\sqrt{p}B}{2}- p\log \frac{1}{4},
\]
a.s., so $\xi^-_n\to\infty$ cannot happen.

Now assume (w.l.o.g.) that there is a compact set ${\cal C}$ so that
for large enough $n$, a.s., $\bxi_{n1},\ldots,\bxi_{n(K-1)}\in {\cal C}$,
but $\|\bxi_{nK}\|\to\infty$. Choose ${\cal C}$ large enough that it also
contains
all components of $\tilde\bxi_{K}$ (from the optimizer $T_K(P)$).

Consider the first term of
$V_{n,K}(\btheta_n,\tilde\bxi_n,\blambda_n,\tilde\x_n):$
\[
W^*_{n,K}(\btheta_n,\tilde\bxi_n,\blambda_n,\tilde\x_n) = \frac{1}{n}\sum_{i=1}^n \min_{k \in \{1,\ldots,K\}} \sum_{j=1}^p \lambda_j \Q(x_{ij},\theta_j,\xi_{kj}).
\]
Define, for any $\x$ and $K$,
\[
C_{nK}(\x)=\argmin_{k\in\{1,\ldots,K\}}\sum_{j=1}^p \lambda_{nj} \Q(x_j,\theta_{nj},\xi_{nkj}),
\]
Then,
\begin{eqnarray*}
W^*_{n,K}(\btheta_n,\tilde\bxi_n,\blambda_n,\tilde\x_n) &=& \frac{1}{n}\sum_{i=1}^n \sum_{j=1}^p \lambda_{nj}
\Q(x_{ij},\theta_{nj},\xi_{nC_{nK}(\x_i)j})\\
&=& \frac{1}{n}\sum_{C_{nK}(\x_n)\neq K}  \sum_{j=1}^p \lambda_{nj}
\Q(x_{ij},\theta_{nj},\xi_{nC_{nK}(\x_i)j})\\
&& + \frac{1}{n}\sum_{C_{nK}(\x_i)= K}
 \sum_{j=1}^p \lambda_{nj}
\Q(x_{ij},\theta_{nj},\xi_{nKj})\\
&\le&  \frac{1}{n}\sum_{C_{nK}(\x_i)\neq K}  \sum_{j=1}^p \lambda_{nj}
\Q(x_{ij},\theta_{nj},\xi_{nC_{nK}(\x_i)j})\\
&&+ \frac{1}{n}\sum_{C_{nK}(\x_i)= K}
 \min_{k\in\{1,\ldots,K-1\}}\sum_{j=1}^p \lambda_{nj}
\Q(x_{ij},\theta_{nj},\xi_{nkj}).
\\
&=& \frac{1}{n}\sum_{i=1}^n \sum_{j=1}^p \lambda_{nj}
\Q(x_{ij},\theta_{nj},\xi_{nC_{n(K-1)}(\x_i)j})\\
&=&
W^*_{n,K-1}(\btheta_n,\tilde\bxi^*_n,\blambda_n,\tilde\x_n),
\end{eqnarray*}
where $\tilde\bxi^*_n=\{\bxi_{n1},\ldots,\bxi_{n(K-1)}\}$.

Consider any set $M=\{\|\x\|\le m\}$ with $m<\infty$.
Observe that, for large enough $n$,
$M\cap \{\x:\ C_{nK}(\x)= K\}=\emptyset$. Furthermore,
\[
\frac{1}{n}\sum_{\indicator{C_{nK}(\x_i)= K}}
\min_{k\in\{1,\ldots,K-1\}}\sum_{j=1}^p \lambda_{nj}
\Q(x_{ij},\theta_{nj},\xi_{nkj})
\le
\frac{1}{n}\sum_{\indicator{C_{nK}(\x_i)= K}}\lambda^+ B_1.
\]
For large enough $n$ this converges, a.s., to $P(\|\x\|>m)\lambda^+ B_1$, which
can be made arbitrarily small by choosing $m$ large enough.

Therefore, for arbitrarily small $\delta>0$ and $n$ large enough,
\begin{eqnarray}
W^*_{n,K}(\btheta_n,\tilde\bxi_n,\blambda_n,\tilde\x_n) &\le&
W^*_{n,K-1}(\btheta_n,\tilde\bxi^*_n,\blambda_n,\tilde\x_n)\nonumber\\
&\le &  \frac{1}{n}\sum_{C_{nK}(\x_i)\neq K} \sum_{j=1}^p
\lambda_{nj} \Q(x_{ij},\theta_{nj},\xi_{nC_{nK}(\x_i)j})+\delta\nonumber\\
&\le & W^*_{n,K}(\btheta_n,\tilde\bxi_n,\blambda_n,\tilde\x_n)+\delta.\label{eq:wk1}
\end{eqnarray}
In order to make use of this, a uniform convergence argument is needed.
Recall $(\btheta_n,\tilde\bxi^*_n,\blambda_n)\in {\cal C}$.
According to \cite{vdvaart98}, Example 19.8 (sometimes referred to as ``uniform law of large numbers''), if ${\cal F}=\{f_{\theta}:\ \theta\in \Theta\}$ is a set of measurable functions with $\theta\mapsto f_\theta(x)$ continuous for all $x$, $\Theta$ compact, and $\exists F\ge |f_\theta| \forall \theta\in \Theta,\ \int F dP<\infty$, then
\[
\sup_{\theta\in\Theta} \left|\frac{1}{n}\sum_{i=1}^n f_\theta(\x_i)-\int f_\theta(\x)dP(\x)\right|\to 0\mbox{ a.s.}
\]
For fixed $x$,
$\Q(x,\theta, \xi)=\left\{\theta+(1-2\theta)\indicator{x<\xi} \right \}|x-\xi|$
is continuous in $(\xi,\theta)$, because
$\xi\to x\Rightarrow \Q(x,\theta, \xi)\to 0$ regardless of whether $\xi$
comes from above or from below. Therefore, for fixed $\x\in\R^p$ and general $K^*$,
\[
U(\btheta,\tilde\bxi,\blambda,\x)=\min_{k \in \{1,\ldots,K^*\}} \sum_{j=1}^p \lambda_j \Q(x_{j},\theta_j,\xi_{kj})-\sum_{j=1}^p\log \left[\lambda_j(\theta_j(1-\theta_j))\right]
\]
is continuous as minimum of continuous functions.

$U(\btheta,\bxi,\blambda,\x)$ can be bounded by a
$P$-integrable function: Let $\xi^+$ an upper bound for the components $|\xi_{kj}|$ (assumed to be in a compact set here). Then,
\begin{eqnarray*}
U(\btheta,\bxi,\blambda,\x)&\le& U^+(\x)=\sum_{j=1}^p \lambda^+(|x_j|+\xi^+) -\sum_{j=1}^p\log \left[\frac{\lambda^-\theta^-}{2}\right],\\
\int U^+(\x) dP(\x) &<&\infty \mbox{ because of A1.}
\end{eqnarray*}
Therefore,
\begin{equation}\label{eq:uloln}
\sup_{(\btheta,\tilde\bxi,\blambda)\in{\cal C}}
|V_{n,K^*}(\btheta,\tilde\bxi,\blambda,\tilde\x_n)-V_{K^*}(\btheta,\tilde\bxi,\blambda,P)|\to 0 \mbox{ a.s.}
\end{equation}
In particular,
\[
\sup_{(\btheta,\tilde\bxi,\blambda)\in{\cal C}}
|V_{n,K-1}(\btheta,\tilde\bxi,\blambda,\tilde\x_n)-V_{K-1}(\btheta,\tilde\bxi,\blambda,P)|\to 0,
\]
Going back to (\ref{eq:wk1}), choose $m$ large
enough that $S_K<S_{K-1}-\delta$. By definition of the optimizers,
\[
V_{n,K}(\btheta_n,\tilde\bxi_n,\blambda_n,\tilde\x_n) \le
V_{n,K}(\btheta_K,\tilde\bxi_K,\blambda_K,\tilde\x_n)\to S_K\mbox{ a.s.,}
\]
and, for large enough $n$, a.s.,
\[
S_{K-1}\le V_{n,K-1}(\btheta_n,\tilde\bxi^*_n,\blambda_n,\tilde\x_n),
\]
but also, eventually,
\[
S_K\ge V_{n,K}(\btheta_n,\tilde\bxi_n,\blambda_n,\tilde\x_n)\ge
V_{n,K-1}(\btheta_n,\tilde\bxi^*_n,\blambda_n,\tilde\x_n)-\delta
\ge S_{K-1}-\delta,
\]
contradicting $S_K < S_{K-1}-\delta$. This implies that
$\bxi_{nk}$ is eventually also captured in a convex set
${\cal C}$.

(\ref{eq:uloln}) now ensures uniform convergence of $V_{n,K}$ to $V_K$
over ${\cal C}$.
The existence of an integrable envelope of $U$ together with
continuity of $U$ imply the continuity of $V_K$
as function of $(\btheta,\tilde\bxi,\blambda)\in{\cal C}$. This and A2 imply
$T_{n,K}(\tilde\x_n)\to T_K(P)$ a.s., because otherwise with probability $>0$
a subsequence of $T_{n,K}(\tilde\x_n)$ can converge against $(\btheta^*,\tilde\bxi^*,\blambda^*)\neq T_K(P)$ but
$\in{\cal C}$ and  with
$V_K(\btheta^*,\tilde\bxi^*,\blambda^*,P)=V_K(T_K(P))$, with contradiction
to A2.

\subsection{Proof of Theorem \ref{tlevel}}\label{proof:tlevel}
The idea here is to show that if for arbitrarily large $m$ a
cluster in $T_K(P_m)$ can be found that has a nonempty intersection with at
least two of the central sets $\{\|\x-\brho_{mk}\|<\epsilon\}$, $S_K$ would be
larger than what could be achieved by putting all the cluster barycenters at
the cluster centers, contradicting the optimality of $T_K(P_m)$.

Write $\gamma_m=\gamma_{T_K(P_m)}$. Define
$(\btheta^*_m,\tilde\bxi^*_m,\blambda^*_m)$ as follows.
For $j=1,\ldots,p,\ k=1,\ldots,K$,
$\theta^*_{mj}=\frac{1}{2},\ \lambda^*_{mj}=1,\ \xi^*_{mkj}=\rho_{mkj}$.
Then, because of A4,
\[
V_K(\btheta^*_m,\tilde\bxi^*_m,\blambda^*_m,P_m)\le \int\sum_{j=1}^p
\frac{1}{2}|x_j-\rho_{mkj}|dP_m(\x)-p\log\frac{1}{4}\le \frac{B}{2}
-p\log\frac{1}{4}.
\]
Similar to the proof of Theorem \ref{tcons},
$\exists \theta^->0, \lambda^->0$ so that for large enough $m$:
$\min(\theta_{m1},\ldots,\theta_{mp},1-\theta_{m1},\ldots,
1-\theta_{mp})\ge \theta^-,$
$\min(\lambda_{m1},\ldots,\lambda_{mp})\ge \lambda^-$, because otherwise the penalty term $-\sum_{j=1}^p\log \lambda_{mj}\theta_{mj}(1-\theta_{mj})$ can be driven to infinity and $(\btheta^*_m,\tilde\bxi^*_m,\blambda^*_m)$ would achieve a smaller and therefore better $V_K$.

For $k_1, k_2\in\{1,\ldots,K\}$
let $I_{mk_1k_2}=\{\|\x-\brho_{mk_1}\|<\epsilon\} \cap \{\gamma_m(\x)=k_2\}$.
Now assume that for at least a subsequence of $m\to\infty$, eventually,
\[
I_{m11}\neq \emptyset \mbox{ and }
I_{m21}\neq \emptyset,
\]
where the cluster numbering has been chosen so that, w.l.o.g.,
\begin{equation}\label{eq:minii}
\min[P_m(I_{m11}),P_m(I_{m21})]=\max_{(k_1, k_2, k_3)\in{\cal K}}
\{\min[P_m(I_{mk_1k_3}),P_m(I_{mk_2k_3})]\}.
\end{equation}
Suppose first that
\[
\limsup_{m\to \infty} \min[P_m(I_{m11}),P_m(I_{m21})]=\tau>0.
\]
Let
$b_m=\max(\|\bxi_{m1}-\brho_{m1}\|-\epsilon,\|\bxi_{m1}-\brho_{m2}\|-\epsilon)$.
Because of A3, $\lim_{m\to\infty}b_m=\infty$. Obviously, for at least one
$k\in\{1,2\}$ and all $\x\in \{\|\x-\brho_{mk}\|<\epsilon\}$:
\[
\sum_{j=1}^p|x_j-\xi_{m1j}|\ge\|\x-\bxi_{m1}\|\ge b_m.
\]
Then
\begin{eqnarray}
V_{K}(\btheta_m,\tilde\bxi_m,\blambda_m,P_m) &\ge & \int \indicator{\gamma_m(\x)=1} \sum_{j=1}^p \lambda_{mj} \Q(x_{j},\theta_j,\xi_{m1j})dP_m(\x)\nonumber\\
&& -\sum_{j=1}^p\log \lambda_{mj}\theta_{mj}(1-\theta_{mj})\nonumber\\
&\ge & \tau\lambda^-\theta^-b_m-p\log\frac{\lambda^+}{4}\to\infty,
\label{eq:wupper}
\end{eqnarray}
so this cannot happen for the minimizer of $V_K$.

Therefore assume w.l.o.g. $\limsup_m P_m(I_{m11})=0$. If also
$\limsup_m P_m(I_{m21})=0$, for this subsequence, $\{\gamma_m(\x)=1\}$
has no nonzero probability overlap with any mixture component's central set
(all of which have probability $\ge \delta$ because of (\ref{eq:a5})),
and there are $K-1$ clusters left to cover $K$ central sets, in contradiction
to (\ref{eq:minii}). Therefore $\limsup_m P_m(I_{m21})=\tau>0$. This means that
$\|\bxi_{m1}-\brho_{m2}\|$ must be bounded, otherwise the argument leading to
(\ref{eq:wupper}) applies again. Therefore $\|\bxi_{m1}-\brho_{m1}\|$ is
unbounded. There must be another cluster,  w.l.o.g., $\{\gamma_m(\x)=2\}$,
so that $P_m(I_{m12})=\tau^*>0$. For $\x\in I_{m11}\neq\emptyset:$
\begin{eqnarray*}
&&\sum_{j=1}^p\lambda_{mj}(\theta_{mj}+(1-2\theta_{mj})\indicator{x_j<\xi_{m1j}})
|x_j-\xi_{m1j}| \\
&\le & \sum_{j=1}^p\lambda_{mj}(\theta_{mj}+(1-2\theta_{mj})\indicator{x_j<\xi_{m2j}})|x_j-\xi_{m2j}|.
\end{eqnarray*}
This, together with
$\|\bxi_{m1}-\brho_{m1}\|\to\infty$ at least for a subsequence, enforces
$\|\bxi_{m2}-\brho_{m1}\|\to\infty$ as well.
But then, as above, with $b^*_m=\|\bxi_{m2}-\brho_{m1}\|-\epsilon$,
\begin{eqnarray*}
V_{K}(\btheta_m,\tilde\bxi_m,\blambda_m,P_m) &\ge & \int \indicator{\gamma_m(\x)=2} \sum_{j=1}^p \lambda_{mj} \Q(x_{j},\theta_j,\xi_{m2j})dP_m(\x)\\
&& -\sum_{j=1}^p\log \lambda_{mj}\theta_{mj}(1-\theta_{mj})\\
&\ge & \tau^*\lambda^-\theta^-b^*_m-p\log\frac{\lambda^+}{4}\to\infty,
\end{eqnarray*}
and again this is impossible for the minimizer of $V_K$.

Taken together, with any numbering of clusters,
\[
I_{m11}\neq \emptyset \mbox{ and }
I_{m21}\neq \emptyset
\]
cannot happen together, so all the central
sets $\{\|\x-\brho_{mk}\|<\epsilon\},\ k\in\{1,\ldots,K\}$ must eventually
be subsets of different clusters.

\newpage
\section{Appendix: Detailed results of the simulation study}
Simulation study on the performance of the quantile-based clustering algorithm. Five different scenarios are considered:
\begin{enumerate}
\item Symmetric Multivariate Student $t$-distributed variables $W \sim t_3$; data come from $K=2,3$ and 5 populations:
    \begin{itemize}
      \item $K=2$, two populations $\mathbf{X}^{(1)}$ and $\mathbf{X}^{(2)}$: $X_j^{(1)} \sim W_j$ and $X_j^{(2)} \sim (
      W_j+1)$, $j=1,\ldots,p$.
      \item $K=3$, three populations $\mathbf{X}^{(1)}$, $\mathbf{X}^{(2)}$ and $\mathbf{X}^{(3)}$: $X_j^{(1)} \sim W_j$, $X_j^{(2)} \sim (W_j+1)$ and $X_j^{(3)} \sim (W_j-1)$, $j=1,\ldots,p$.
      \item $K=5$, five populations $\mathbf{X}^{(1)}$, $\mathbf{X}^{(2)}$, $\mathbf{X}^{(3)}$, $\mathbf{X}^{(4)}$ and $\mathbf{X}^{(5)}$: $X_j^{(1)} \sim W_j$, $X_j^{(2)} \sim (W_j+1)$, $X_j^{(3)} \sim (W_j+2)$, $X_j^{(4)} \sim (W_j-1)$ and $X_j^{(5)} \sim (W_j-2)$, $j=1,\ldots,p$.
    \end{itemize}

\item Highly skewed data i.i.d. vectors $W \sim MVN(\mathbf{0}_p, \mathbf{\Sigma})$ transformed by
using the exponential function; data come from $K=2,3$ and 5 populations:
    \begin{itemize}
      \item $K=2$, two populations $\mathbf{X}^{(1)}$ and $\mathbf{X}^{(2)}$: $X_j^{(1)} \sim \exp(W_j)$ and $X_j^{(2)} \sim (\exp(W_j)+0.6)$, $j=1,\ldots,p$.
      \item $K=3$, three populations $\mathbf{X}^{(1)}$, $\mathbf{X}^{(2)}$ and $\mathbf{X}^{(3)}$: $X_j^{(1)} \sim \exp(W_j)$, $X_j^{(2)} \sim (\exp(W_j)+0.6)$ and $X_j^{(3)} \sim (\exp(W_j)-0.6)$, $j=1,\ldots,p$.
      \item $K=5$, five populations $\mathbf{X}^{(1)}$, $\mathbf{X}^{(2)}$, $\mathbf{X}^{(3)}$, $\mathbf{X}^{(4)}$ and $\mathbf{X}^{(5)}$: $X_j^{(1)} \sim \exp(W_j)$, $X_j^{(2)} \sim (\exp(W_j)+0.6)$, $X_j^{(3)} \sim (\exp(W_j)+1.2)$, $X_j^{(4)} \sim (\exp(W_j)-0.6)$ and $X_j^{(5)} \sim (\exp(W_j)-1.2)$, $j=1,\ldots,p$.
    \end{itemize}

\item Different distributions for the $p$ variables. Firstly, $W \sim MVN(\mathbf{0}_p, \mathbf{\Sigma})$ and then split $p$ into five balanced blocks to which different transformation were applied; data come from $K=2,3$ and 5 populations (subscripts in square brackets indicate variable block):
    \begin{itemize}
      \item $K=2$, two populations $\mathbf{X}^{(1)}$ and $\mathbf{X}^{(2)}$:
            \begin{itemize}
        		\item $X_{j[1]}^{(1)} \sim W_j$ and $X_{j[1]}^{(2)} \sim W_j+0.7$, $j=1,\ldots,p$;
        		\item $X_{j[2]}^{(1)} \sim exp(W_j)$ and $X_{j[2]}^{(2)} \sim exp(W_j+0.7)$, $j=1,\ldots,p$.;
        		\item $X_{j[3]}^{(1)} \sim log(|W_j|)$  and $X_{j[3]}^{(2)}  \sim  log(|W_j +0.7|)$, $j=1,\ldots,p$;
        		\item $X_{j[4]}^{(1)} \sim W_j^2$ and $X_{j[4]}^{2} \sim (W_j+0.7)^2$, $j=1,\ldots,p$;
        		\item $X_{j[5]}^{(1)} \sim \sqrt{|W_j|}$ and $X_{j[5]}^{(2)} \sim \sqrt{|W_j+0.7|}$, $j=1,\ldots,p$.
        	\end{itemize}
      \item $K=3$, three populations $\mathbf{X}^{(1)}$, $\mathbf{X}^{(2)}$ and $\mathbf{X}^{(3)}$:
            \begin{itemize}
        		\item $X_{j[1]}^{(1)} \sim W_j$, $X_{j[1]}^{(2)} \sim W_j+0.7$ and $X_{j[1]}^{(3)} \sim (W_j+1.4)$, $j=1,\ldots,p$;
        		\item $X_{j[2]}^{(1)} \sim exp(W_j)$, $X_{j[2]}^{(2)} \sim exp(W_j+0.7)$ and $X_{j[2]}^{(3)} \sim exp(W_j+1.4)$, $j=1,\ldots,p$;
        		\item $X_{j[3]}^{(1)} \sim log(|W_j|)$, $X_{j[3]}^{(2)} \sim  log(|W_j +0.7|)$ and $X_{j[3]}^{(3)} \sim  log(|W_j +1.4|)$, $j=1,\ldots,p$;
        		\item $X_{j[4]}^{(1)} \sim W_j^2$, $X_{j[4]}^{2} \sim (W_j+0.7)^2$ and $X_{j[4]}^{3} \sim (W_j+1.4)^2$, $j=1,\ldots,p$;
        		\item $X_{j[5]}^{(1)} \sim \sqrt{|W_j|}$, $X_{j[5]}^{(2)} \sim \sqrt{|W_j+0.7|}$ and $X_{j[5]}^{(3)} \sim \sqrt{|W_j+1.4|}$, $j=1,\ldots,p$.
        	\end{itemize}
      \item $K=5$, five populations $\mathbf{X}^{(1)}$, $\mathbf{X}^{(2)}$, $\mathbf{X}^{(3)}$, $\mathbf{X}^{(4)}$ and $\mathbf{X}^{(5)}$:
            \begin{itemize}
        		\item $X_{j[1]}^{(1)} \sim W_j$, $X_{j[1]}^{(2)} \sim W_j+0.7$, $X_{j[1]}^{(3)} \sim (W_j+1.4)$, $X_{j[1]}^{(4)} \sim (W_j+2.1)$ and $X_{j[1]}^{(5)} \sim (W_j+2.8)$, $j=1,\ldots,p$;
        		\item $X_{j[2]}^{(1)} \sim exp(W_j)$, $X_{j[2]}^{(2)} \sim exp(W_j+0.7)$, $X_{j[2]}^{(3)} \sim exp(W_j+1.4)$, $X_{j[2]}^{(4)} \sim exp(W_j+2.1)$ and $X_{j[2]}^{(5)} \sim exp(W_j+2.8)$, $j=1,\ldots,p$;
        		\item $X_{j[3]}^{(1)} \sim log(|W_j|)$, $X_{j[3]}^{(2)} \sim  log(|W_j +0.7|)$, $X_{j[3]}^{(3)} \sim  log(|W_j +1.4|)$, $X_{j[3]}^{(4)} \sim  log(|W_j +2.1|)$ and $X_{j[3]}^{(5)} \sim  log(|W_j +2.8|)$, $j=1,\ldots,p$;
        		\item $X_{j[4]}^{(1)} \sim W_j^2$, $X_{j[4]}^{(2)} \sim (W_j+0.7)^2$, $X_{j[4]}^{(3)} \sim (W_j+1.4)^2$, $X_{j[4]}^{(4)} \sim (W_j+2.1)^2$ and $X_{j[4]}^{(5)} \sim (W_j+2.8)^2$ , $j=1,\ldots,p$;
        		\item $X_{j[5]}^{(1)} \sim \sqrt{|W_j|}$, $X_{j[5]}^{(2)} \sim \sqrt{|W_j+0.7|}$, $X_{j[5]}^{(3)} \sim \sqrt{|W_j+1.4|}$, $X_{j[5]}^{(4)} \sim \sqrt{|W_j+2.1|}$ and $X_{j[5]}^{(5)} \sim \sqrt{|W_j+2.8|}$, $j=1,\ldots,p$.
        	\end{itemize}
    \end{itemize}
\item Different distributional shapes and levels of skewness even for different classes within the same variable. Within each class, data were generated according to beta distributions with parameters $a$ and $b$ in the interval $(1,10)$
    randomly generated for each class within each variable. The absolute difference between the class expected values is bounded from above by 0.2:
    \begin{itemize}
          \item $K=2$, two populations $\mathbf{X}^{(1)}$ and $\mathbf{X}^{(2)}$: $X_{j}^{(1)} \sim \mbox{Beta}(\alpha,\beta)$, where $\alpha, \beta \sim U(1,10)$, and $X_{j}^{(2)} \sim \mbox{Beta}(\alpha,\beta)$, where $\alpha, \beta \sim U(1,10)$, $j=1,\ldots,p$.
      \item $K=3$, three populations $\mathbf{X}^{(1)}$, $\mathbf{X}^{(2)}$ and $\mathbf{X}^{(3)}$: $X_{j}^{(1)} \sim \mbox{Beta}(\alpha,\beta)$, where $\alpha, \beta \sim U(1,10)$, $X_{j}^{(2)} \sim \mbox{Beta}(\alpha,\beta)$, where $\alpha, \beta \sim U(1,10)$, and $X_{j}^{(3)} \sim \mbox{Beta}(\alpha,\beta)$, where $\alpha, \beta \sim U(1,10)$, $j=1,\ldots,p$.
      \item $K=5$, five populations $\mathbf{X}^{(1)}$, $\mathbf{X}^{(2)}$, $\mathbf{X}^{(3)}$, $\mathbf{X}^{(4)}$ and $\mathbf{X}^{(5)}$: $X_{j}^{(1)} \sim \mbox{Beta}(\alpha,\beta)$, where $\alpha, \beta \sim U(1,10)$, $X_{j}^{(2)} \sim \mbox{Beta}(\alpha,\beta)$, where $\alpha, \beta \sim U(1,10)$, $X_{j}^{(3)} \sim \mbox{Beta}(\alpha,\beta)$, where $\alpha,\beta \sim U(1,10)$, $X_{j}^{(4)} \sim \mbox{Beta}(\alpha,\beta)$, where $\alpha, \beta \sim U(1,10)$, and $X_{j}^{(5)} \sim \mbox{Beta}(\alpha,\beta)$, where $\alpha,\beta \sim U(1,10)$, $j=1,\ldots,p$.
    \end{itemize}

\item Different distributional shapes and
levels of skewness even for different classes within each variable. Within each class, data
are generated according to beta distributions with parameters $a$ and $b$ randomly chosen to be in the intervals: $(0,1)$ and $(1,5)$, or  $(0,1)$ and $(1,5)$, $(1,3)$ and $(5,10)$,$(1,3)$ and $(1,3)$, for each class within each variable. The absolute difference between the class expected values is bounded from above by 0.1, and the so chosen interval guarantees a higher level of skewness for some variables.
    \begin{itemize}
          \item $K=2$, two populations $\mathbf{X}^{(1)}$ and $\mathbf{X}^{(2)}$: $X_{j}^{(1)} \sim \mbox{Beta}(\alpha,\beta)$, where either:
               \begin{itemize}
               \item $\alpha \sim U(0.1,1)$ and $\beta \sim U(1,10)$, or
               \item $\alpha \sim U(1,10)$ and $\beta \sim U(0.1,1)$;
               \end{itemize}
               and the same for $X_{j}^{(2)}$, $j=1,\ldots,p$.
      \item $K=3$, three populations $\mathbf{X}^{(1)}$, $\mathbf{X}^{(2)}$ and $\mathbf{X}^{(3)}$: $X_{j}^{(1)} \sim \mbox{Beta}(\alpha,\beta)$, where either:
                \begin{itemize}
                  \item $\alpha \sim U(0.1,1)$ and $\beta \sim U(1,10)$, or
                  \item $\alpha \sim U(1,10)$ and $\beta \sim U(0.1,1)$;
                \end{itemize}
                 and the same for $X_{j}^{(2)}$; $X_{j}^{(3)} \sim Beta(\alpha,\beta)$, where $\alpha \sim U(1,3)$ and $\beta \sim U(5,10)$,  $j=1,\ldots,p$.
      \item $K=5$, five populations $\mathbf{X}^{(1)}$, $\mathbf{X}^{(2)}$, $\mathbf{X}^{(3)}$, $\mathbf{X}^{(4)}$ and $\mathbf{X}^{(5)}$: $X_{j}^{(1)} \sim \mbox{Beta}(\alpha,\beta)$, where either
           \begin{itemize}
                  \item $\alpha \sim U(0.1,1)$ and $\beta \sim U(1,5)$, or
                  \item $\alpha \sim U(1,5)$ and $\beta \sim U(0.1,1)$, or
                  \item $\alpha \sim U(1,3)$ and $\beta \sim U(5,10)$, or
                  \item $\alpha \sim U(5,10)$ and $\beta \sim U(1,3)$, or
                  \item $\alpha \sim U(1,3)$ and $\beta \sim U(1,3)$;

           \end{itemize}
            and the same for $X_{j}^{(2)}$, $X_{j}^{(3)}$, $X_{j}^{(4)}$ an $X_{j}^{(5)}$ $j=1,\ldots,p$.
    \end{itemize}

\end{enumerate}

For each of the five scenarios and for each set of $K$ populations, $K=2,3,5$, we evaluated combinations of $p=\{50, 100, 500\}$, $n=\{50,100, 500\}$, different percentages of relevant variables for grouping structure, i.e., $100\%$, $50\%$ and $10\%$,
independent or dependent variables (limitedly to scenario (a)-(c)), for a total of 648 different settings. The dependence structure
among variables was modeled via the function \texttt{rcorrmatrix} from the R package \texttt{clusterGeneration}, so that the correlation matrices are uniformly distributed over the space of positive definite
correlation matrices, with each correlation marginally distributed as $Beta(p/2, p/2)$ on
$(-1, 1)$. The irrelevant noise variables were generated independently of each other from the base distribution of each scenario.

The number of clusters is taken as known (and equal to the number of populations data are generated from) for every method.
The clustering procedures that have been considered are the following:

\begin{itemize}
\item Common $\theta$ and Unscaled variables (CU) $k$-quantile clustering algorithm;
\item Variable-wise $\theta_j$ and Unscaled variables (VU) $k$-quantile clustering algorithm;
\item Common $\theta$ and Scaled variables (CS) $k$-quantile clustering algorithm;
\item Variable-wise $\theta_j$ and Scaled variables (VS) $k$-quantile clustering algorithm;
 \item Mixture of Gaussian distributions, estimated by the default options of function \texttt{Mclust} from the R package \texttt{mclust};
\item Mixture of skew-$t$ distributions, estimated by the \texttt{EmSkew} function from the R package \texttt{EMMIXskew}, argument \texttt{distr} equal to \texttt{mst}, and initialised by the $k$-means clustering algorithm;
\item Mixture of Factor Analyzers, estimated by the \texttt{fma} function from the \texttt{FactMixtAnalysis} R package, by fitting models with number of latent factors from 1 to 20;
\item $k$-means clustering algorithm, run by the \texttt{kmeans} function from the \texttt{stats} R package, with five random starts;
 \item Partition Around Medoids, run by the default options of the \texttt{pam} function from the \texttt{cluster} R package;
\item Agglomerative hierarchical clustering with average link, run by the \texttt{hclust} function, option \texttt{method='average'}, of the \texttt{stats} R package;
\item Spectral clustering, estimated by the default options of function \texttt{specc} from the R package \texttt{kernlab};
\item Affinity Propagation clustering, estimated by the function \texttt{apclusterK}, with similarities computed as squared negative distances, from the  R package \texttt{apcluster}.
\end{itemize}

For each setting 100 simulations were run. The average Adjusted Rand Index values and the corresponding standard errors are reported in the following tables, multiplied by 100; for each method and scenario the number of valid cases out of 100 runs is included as well.

Table \ref{tab:SimTimes} contains the average times (in seconds) - and the corresponding standard errors in brackets - required by each algorithm (excluding mixtures of $t$s, mixtures of skew-Normals and mixtures of skew-$t$s, as they could not always reach the convergence) to cluster the a single data set from each of the five scenarios, considering the cases with 50\%  of relevant variables, $K=2$, both dependent and independent variables; all the procedures run on a Lenovo PC, Intel Core i5-6500 CPU, 3.20 GHz, 20 Gb of RAM. NaN/NA values mean that the method did not deliver a solution for at least one dataset.

\bigskip

We further study the behaviour of the algorithm in characterizing the variables through the estimation of $theta$ and $lambda$.
In particular, we consider the setting of scenario 3, where features are not all identically distributed, for the cases of $K=3$, $n=500$ and $p=50$, independent of each other. The boxplots in the Figure below show the distribution of the estimated parameters for the variables grouped according to their distribution. As it can be seen, $\theta$ captures the skewness of the data: the estimated parameters are about 0.5 when data come from a symmetric (Gaussian) distribution, between 0 and 0.5 with positively skewed data and between 0.5 and 1 for negatively skewed data. Differently, the parameter $\lambda$ mainly accounts for the scaling, but its characterization is less specific; the variability of the results is due not only to the aggregation of 100 replicates but also of three clusters for which potentially different parameters are optimal.

\begin{center}
\includegraphics[width=0.75\textwidth]{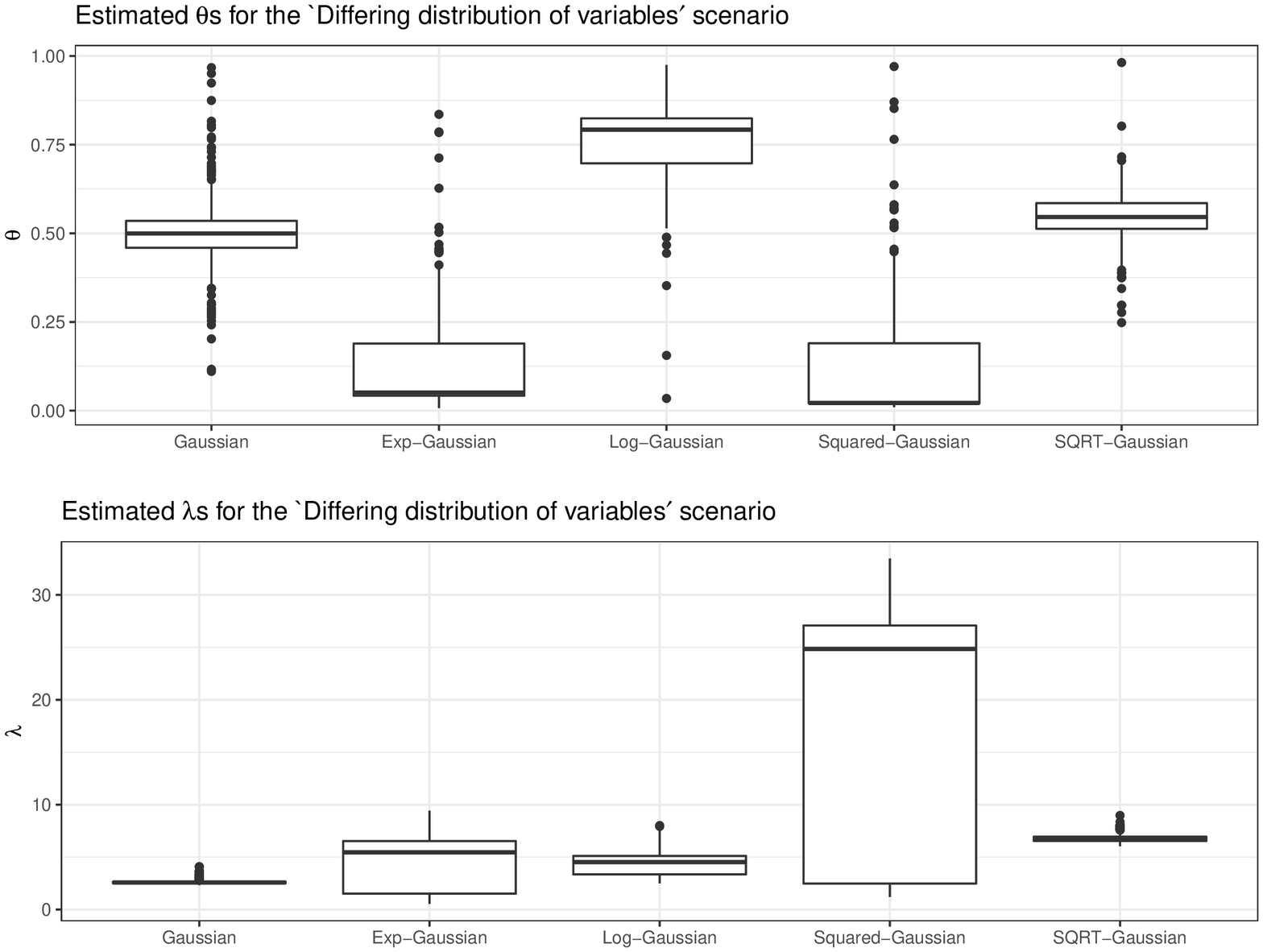}
\end{center}

\newgeometry{a4paper,
 left=20mm,
 top=20mm,bottom=20mm
}

\begin{sidewaystable}
{\textwidth=3\linewidth
\caption{Simulation study: independent identically distributed symmetric variables, $K=2$ groups. Average Adjusted Rand index $\times 100$ (with standard errors $\times 100$, in brackets) for different methods. Numbers in squared brackets indicate the valid cases out of 100 runs. \aftergroup\aftergroup\aftergroup
\aftergroup\aftergroup\aftergroup
\aftergroup
\leavevmode}}
\centering \footnotesize
\begin{tabular}{lrrrrrrrrr}
\hline
 & \multicolumn{9}{c}{$n=50$}    \\
  \hline
  & \multicolumn{3}{c}{$p=50$} & \multicolumn{3}{c}{$p=100$} & \multicolumn{3}{c}{$p=500$} \\
   \hline
 & $100\%$ & $50\%$ &    $10\%$ & $100\%$ & $50\%$  & $10\%$ & $100\%$ & $50\%$  & $10\%$ \\
CU & 88.32 $(0.98)_{[100]}$ & 73.01 $(1.28)_{[100]}$ & 8.77 $(0.93)_{[100]}$ & 94.48 $(0.65)_{[100]}$ & 84.01 $(1.09)_{[100]}$ & 16.63 $(1.39)_{[100]}$ & 99.36 $(0.22)_{[100]}$ & 96.98 $(0.46)_{[100]}$ & 38.31 $(1.49)_{[100]}$ \\
  VU & 87.90 $(0.93)_{[100]}$ & 72.94 $(1.47)_{[100]}$ & 6.71 $(0.94)_{[100]}$ & 94.38 $(0.60)_{[100]}$ & 83.55 $(1.09)_{[100]}$ & 14.05 $(1.20)_{[100]}$ & 99.12 $(0.27)_{[100]}$ & 96.45 $(0.58)_{[100]}$ & 39.49 $(1.70)_{[100]}$ \\
  CS & 88.20 $(0.91)_{[100]}$ & 72.72 $(1.27)_{[100]}$ & 8.21 $(0.90)_{[100]}$ & 94.48 $(0.65)_{[100]}$ & 83.71 $(1.09)_{[100]}$ & 16.15 $(1.40)_{[100]}$ & 99.36 $(0.22)_{[100]}$ & 96.98 $(0.46)_{[100]}$ & 38.30 $(1.49)_{[100]}$ \\
  VS & 85.91 $(1.03)_{[100]}$ & 69.93 $(1.57)_{[100]}$ & 5.86 $(0.83)_{[100]}$ & 91.56 $(1.08)_{[100]}$ & 79.33 $(1.59)_{[100]}$ & 12.04 $(1.43)_{[100]}$ & 94.06 $(1.28)_{[100]}$ & 86.91 $(2.36)_{[100]}$ & 28.03 $(2.00)_{[100]}$ \\
  Mixt-N & 18.52 $(3.52)_{[100]}$ & 1.81 $(1.09)_{[100]}$ & 0.04 $(0.05)_{[100]}$ & 24.09 $(4.02)_{[100]}$ & 5.90 $(2.20)_{[100]}$ & -0.05 $(0.03)_{[100]}$ & 30.31 $(4.48)_{[100]}$ & 4.10 $(1.97)_{[100]}$ & -0.03 $(0.03)_{[100]}$ \\
\hline
\end{tabular}
\end{sidewaystable}

 \begin{sidewaystable}
{\textwidth=3\linewidth
\caption{\label{t:sim1b}Simulation study: dependent identically distributed symmetric variables, $K=2$ groups.  Average Adjusted Rand index $\times 100$ (with standard errors $\times 100$, in brackets) for different methods. Numbers in squared brackets indicate the valid cases out of 100 runs. \aftergroup\aftergroup\aftergroup
\aftergroup\aftergroup\aftergroup
\aftergroup
\leavevmode}}
\centering \footnotesize
\begin{tabular}{lrrrrrrrrr}
\hline
 & \multicolumn{9}{c}{$n=50$}    \\
  \hline
  & \multicolumn{3}{c}{$p=50$} & \multicolumn{3}{c}{$p=100$} & \multicolumn{3}{c}{$p=500$} \\
   \hline
 & $100\%$ & $50\%$ &    $10\%$ & $100\%$ & $50\%$  & $10\%$ & $100\%$ & $50\%$  & $10\%$ \\
 CU & 85.18 $(1.14)_{[100]}$ & 64.91 $(1.46)_{[100]}$ & 4.95 $(0.72)_{[100]}$ & 94.88 $(0.66)_{[100]}$ & 84.94 $(1.02)_{[100]}$ & 12.06 $(1.04)_{[100]}$ & 99.36 $(0.22)_{[100]}$ & 96.74 $(0.48)_{[100]}$ & 38.96 $(1.47)_{[100]}$ \\
  VU & 83.93 $(1.09)_{[100]}$ & 63.48 $(1.69)_{[100]}$ & 3.12 $(0.65)_{[100]}$ & 93.96 $(0.73)_{[100]}$ & 84.76 $(1.00)_{[100]}$ & 9.87 $(1.05)_{[100]}$ & 99.04 $(0.28)_{[100]}$ & 97.07 $(0.50)_{[100]}$ & 39.27 $(1.56)_{[100]}$ \\
  CS & 85.31 $(1.11)_{[100]}$ & 64.36 $(1.49)_{[100]}$ & 4.77 $(0.67)_{[100]}$ & 94.64 $(0.67)_{[100]}$ & 84.64 $(1.02)_{[100]}$ & 11.27 $(1.00)_{[100]}$ & 99.36 $(0.22)_{[100]}$ & 96.66 $(0.48)_{[100]}$ & 38.79 $(1.46)_{[100]}$ \\
  VS & 82.01 $(1.20)_{[100]}$ & 60.70 $(1.68)_{[100]}$ & 3.30 $(0.63)_{[100]}$ & 90.50 $(1.55)_{[100]}$ & 78.45 $(1.81)_{[100]}$ & 7.13 $(1.01)_{[100]}$ & 94.83 $(1.24)_{[100]}$ & 88.50 $(2.36)_{[100]}$ & 30.37 $(2.03)_{[100]}$ \\
  Mixt-N & 18.51 $(3.42)_{[100]}$ & 0.49 $(0.37)_{[100]}$ & -0.02 $(0.05)_{[100]}$ & 26.91 $(4.07)_{[100]}$ & 3.44 $(1.68)_{[100]}$ & 0.12 $(0.15)_{[100]}$ & 31.10 $(4.50)_{[100]}$ & 3.04 $(1.71)_{[100]}$ & -0.03 $(0.02)_{[100]}$ \\
\hline
\end{tabular}
\end{sidewaystable}

\begin{sidewaystable}
{\textwidth=3\linewidth\caption{\label{t:sim2a}Simulation study: independent identically distributed asymmetric variables, $K=2$ groups. Average Adjusted Rand index $\times 100$ (with standard errors $\times 100$, in brackets) for different methods. Numbers in squared brackets indicate the valid cases out of 100 runs.\aftergroup\aftergroup\aftergroup
\aftergroup\aftergroup\aftergroup
\aftergroup
\leavevmode}}
\centering \footnotesize
\begin{tabular}{lrrrrrrrrr}
\hline
 & \multicolumn{9}{c}{$n=50$}    \\
  \hline
  & \multicolumn{3}{c}{$p=50$} & \multicolumn{3}{c}{$p=100$} & \multicolumn{3}{c}{$p=500$} \\
   \hline
 & $100\%$ & $50\%$ &    $10\%$ & $100\%$ & $50\%$  & $10\%$ & $100\%$ & $50\%$  & $10\%$ \\
  CU & 92.33 $(1.52)_{[100]}$ & 10.95 $(1.00)_{[100]}$ & 0.15 $(0.07)_{[100]}$ & 99.92 $(0.08)_{[100]}$ & 3.46 $(0.39)_{[100]}$ & 0.07 $(0.05)_{[100]}$ & 100.00 $(0.00)_{[100]}$ & 0.41 $(0.13)_{[100]}$ & 0.00 $(0.00)_{[100]}$ \\
  VU & 92.38 $(0.90)_{[100]}$ & 46.71 $(2.33)_{[100]}$ & 2.28 $(0.49)_{[100]}$ & 98.25 $(0.37)_{[100]}$ & 74.62 $(1.68)_{[100]}$ & 2.56 $(0.51)_{[100]}$ & 100.00 $(0.00)_{[100]}$ & 99.04 $(0.28)_{[100]}$ & 11.92 $(1.05)_{[100]}$ \\
  CS & 100.00 $(0.00)_{[100]}$ & 48.81 $(4.50)_{[100]}$ & 0.12 $(0.06)_{[100]}$ & 100.00 $(0.00)_{[100]}$ & 13.54 $(3.07)_{[100]}$ & 0.02 $(0.02)_{[100]}$ & 100.00 $(0.00)_{[100]}$ & 0.41 $(0.13)_{[100]}$ & 0.00 $(0.00)_{[100]}$ \\
  VS & 94.50 $(1.66)_{[100]}$ & 74.39 $(3.26)_{[100]}$ & 2.54 $(0.54)_{[100]}$ & 95.03 $(1.53)_{[100]}$ & 83.93 $(2.83)_{[100]}$ & 1.98 $(0.52)_{[100]}$ & 90.33 $(2.66)_{[100]}$ & 80.89 $(3.35)_{[100]}$ & 5.87 $(0.81)_{[100]}$ \\
  Mixt-N & 0.16 $(0.09)_{[100]}$ & -0.01 $(0.04)_{[100]}$ & 0.00 $(0.06)_{[100]}$ & 0.01 $(0.04)_{[100]}$ & -0.06 $(0.02)_{[100]}$ & -0.06 $(0.02)_{[100]}$ & 5.42 $(2.00)_{[100]}$ & 0.00 $(0.01)_{[100]}$ & 0.00 $(0.01)_{[100]}$\\
\hline
\end{tabular}
\end{sidewaystable}

\begin{sidewaystable}
{\textwidth=3\linewidth\caption{\label{t:sim2b}
Simulation study: dependent identically distributed asymmetric variables, $K=2$ groups.  Average Adjusted Rand index $\times 100$ (with standard errors $\times 100$, in brackets) for different methods. Numbers in squared brackets indicate the valid cases out of 100 runs. \aftergroup\aftergroup\aftergroup
\aftergroup\aftergroup\aftergroup
\aftergroup
\leavevmode}}
\centering \footnotesize
\begin{tabular}{lrrrrrrrrr}
\hline
 & \multicolumn{9}{c}{$n=50$}    \\
  \hline
  & \multicolumn{3}{c}{$p=50$} & \multicolumn{3}{c}{$p=100$} & \multicolumn{3}{c}{$p=500$} \\
   \hline
 & $100\%$ & $50\%$ &    $10\%$ & $100\%$ & $50\%$  & $10\%$ & $100\%$ & $50\%$  & $10\%$ \\
  CU & 83.25 $(2.05)_{[100]}$ & 9.75 $(0.78)_{[100]}$ & 0.48 $(0.17)_{[100]}$ & 99.44 $(0.21)_{[100]}$ & 4.30 $(0.54)_{[100]}$ & 0.02 $(0.02)_{[100]}$ & 100.00 $(0.00)_{[100]}$ & 1.23 $(1.00)_{[100]}$ & 0.00 $(0.00)_{[100]}$ \\
  VU & 87.57 $(1.13)_{[100]}$ & 26.95 $(2.15)_{[100]}$ & 1.14 $(0.37)_{[100]}$ & 98.41 $(0.36)_{[100]}$ & 69.17 $(2.19)_{[100]}$ & 1.61 $(0.44)_{[100]}$ & 100.00 $(0.00)_{[100]}$ & 99.28 $(0.25)_{[100]}$ & 12.41 $(1.11)_{[100]}$ \\
  CS & 100.00 $(0.00)_{[100]}$ & 37.64 $(4.17)_{[100]}$ & 0.38 $(0.14)_{[100]}$ & 100.00 $(0.00)_{[100]}$ & 14.98 $(3.15)_{[100]}$ & 0.03 $(0.03)_{[100]}$ & 100.00 $(0.00)_{[100]}$ & 1.24 $(1.00)_{[100]}$ & 0.00 $(0.00)_{[100]}$ \\
  VS & 91.39 $(2.02)_{[100]}$ & 59.00 $(4.04)_{[100]}$ & 1.81 $(0.53)_{[100]}$ & 94.41 $(1.78)_{[100]}$ & 83.46 $(2.60)_{[100]}$ & 1.78 $(0.58)_{[100]}$ & 89.09 $(2.91)_{[100]}$ & 84.24 $(3.18)_{[100]}$ & 6.58 $(0.96)_{[100]}$ \\
  Mixt-N & 0.26 $(0.17)_{[100]}$ & 0.13 $(0.15)_{[100]}$ & -0.06 $(0.03)_{[100]}$ & -0.01 $(0.03)_{[100]}$ & 0.06 $(0.06)_{[100]}$ & -0.03 $(0.02)_{[100]}$ & 3.55 $(1.44)_{[100]}$ & 0.02 $(0.02)_{[100]}$ & 0.01 $(0.02)_{[100]}$ \\
\hline
\end{tabular}
 \end{sidewaystable}

\begin{sidewaystable}
{\textwidth=3\linewidth \caption{\label{t:sim3a}Simulation study: independent not identically distributed variables, $K=2$ groups. Average Adjusted Rand index $\times 100$ (with standard errors $\times 100$, in brackets) for different methods. Numbers in squared brackets indicate the valid cases out of 100 runs. \aftergroup\aftergroup\aftergroup
\aftergroup\aftergroup\aftergroup
\aftergroup
\leavevmode}}
\centering \footnotesize
\begin{tabular}{lrrrrrrrrr}
\hline
 & \multicolumn{9}{c}{$n=50$}    \\
  \hline
  & \multicolumn{3}{c}{$p=50$} & \multicolumn{3}{c}{$p=100$} & \multicolumn{3}{c}{$p=500$} \\
   \hline
 & $100\%$ & $50\%$ &    $10\%$ & $100\%$ & $50\%$  & $10\%$ & $100\%$ & $50\%$  & $10\%$ \\
CU & 55.81 $(1.51)_{[100]}$ & 27.58 $(1.39)_{[100]}$ & 2.11 $(0.48)_{[100]}$ & 72.14 $(1.42)_{[100]}$ & 33.37 $(1.60)_{[100]}$ & 1.33 $(0.34)_{[100]}$ & 97.11 $(0.63)_{[100]}$ & 34.12 $(2.18)_{[100]}$ & 0.24 $(0.07)_{[100]}$ \\
  VU & 55.44 $(1.61)_{[100]}$ & 24.26 $(1.42)_{[100]}$ & 2.39 $(0.48)_{[100]}$ & 84.88 $(1.24)_{[100]}$ & 49.58 $(1.77)_{[100]}$ & 2.85 $(0.58)_{[100]}$ & 100.00 $(0.00)_{[100]}$ & 95.96 $(0.57)_{[100]}$ & 15.28 $(0.98)_{[100]}$ \\
  CS & 54.98 $(1.64)_{[100]}$ & 23.82 $(1.24)_{[100]}$ & 1.71 $(0.48)_{[100]}$ & 72.44 $(1.45)_{[100]}$ & 32.71 $(1.55)_{[100]}$ & 1.22 $(0.32)_{[100]}$ & 97.11 $(0.62)_{[100]}$ & 34.00 $(2.19)_{[100]}$ & 0.21 $(0.07)_{[100]}$ \\
  VS & 51.05 $(1.96)_{[100]}$ & 19.91 $(1.36)_{[100]}$ & 0.82 $(0.33)_{[100]}$ & 69.87 $(2.38)_{[100]}$ & 34.31 $(2.07)_{[100]}$ & 1.80 $(0.47)_{[100]}$ & 79.96 $(3.31)_{[100]}$ & 66.50 $(3.48)_{[100]}$ & 8.63 $(1.04)_{[100]}$ \\
  Mixt-N & 2.22 $(0.64)_{[100]}$ & 0.58 $(0.17)_{[100]}$ & 0.28 $(0.13)_{[100]}$ & 3.41 $(1.34)_{[100]}$ & 2.13 $(0.84)_{[100]}$ & 0.10 $(0.06)_{[100]}$ & 32.15 $(4.35)_{[100]}$ & 8.77 $(2.28)_{[100]}$ & 0.13 $(0.10)_{[100]}$ \\
\hline
\end{tabular}
 \end{sidewaystable}

\begin{sidewaystable}
{\textwidth=3\linewidth \caption{\label{t:sim3b} Simulation study: dependent not identically distributed variables, $K=2$ groups. Average Adjusted Rand index $\times 100$ (with standard errors $\times 100$, in brackets) for different methods. Numbers in squared brackets indicate the valid cases out of 100 runs. \aftergroup\aftergroup\aftergroup
\aftergroup\aftergroup\aftergroup
\aftergroup
\leavevmode}}
\centering \footnotesize
\begin{tabular}{lrrrrrrrrr}
\hline
 & \multicolumn{9}{c}{$n=50$}    \\
  \hline
  & \multicolumn{3}{c}{$p=50$} & \multicolumn{3}{c}{$p=100$} & \multicolumn{3}{c}{$p=500$} \\
   \hline
 & $100\%$ & $50\%$ &    $10\%$ & $100\%$ & $50\%$  & $10\%$ & $100\%$ & $50\%$  & $10\%$ \\
CU & 52.29 $(1.71)_{[100]}$ & 22.12 $(1.41)_{[100]}$ & 1.82 $(0.46)_{[100]}$ & 70.03 $(1.36)_{[100]}$ & 29.33 $(1.47)_{[100]}$ & 1.38 $(0.34)_{[100]}$ & 97.86 $(0.46)_{[100]}$ & 34.82 $(2.06)_{[100]}$ & 0.39 $(0.19)_{[100]}$ \\
  VU & 52.91 $(1.84)_{[100]}$ & 20.74 $(1.47)_{[100]}$ & 2.42 $(0.52)_{[100]}$ & 82.48 $(1.23)_{[100]}$ & 43.56 $(1.72)_{[100]}$ & 2.57 $(0.52)_{[100]}$ & 100.00 $(0.00)_{[100]}$ & 94.18 $(0.70)_{[100]}$ & 14.81 $(1.18)_{[100]}$ \\
  CS & 47.65 $(1.58)_{[100]}$ & 20.77 $(1.29)_{[100]}$ & 1.54 $(0.46)_{[100]}$ & 70.81 $(1.39)_{[100]}$ & 28.61 $(1.40)_{[100]}$ & 1.45 $(0.36)_{[100]}$ & 97.94 $(0.46)_{[100]}$ & 34.81 $(2.08)_{[100]}$ & 0.38 $(0.19)_{[100]}$ \\
  VS & 45.63 $(1.97)_{[100]}$ & 16.59 $(1.36)_{[100]}$ & 1.04 $(0.42)_{[100]}$ & 69.33 $(2.41)_{[100]}$ & 32.17 $(2.01)_{[100]}$ & 1.50 $(0.45)_{[100]}$ & 83.95 $(3.01)_{[100]}$ & 61.43 $(3.29)_{[100]}$ & 8.78 $(1.15)_{[100]}$ \\
  Mixt-N & 24.72 $(2.19)_{[100]}$ & 4.80 $(0.79)_{[100]}$ & 0.51 $(0.28)_{[100]}$ & 48.63 $(2.32)_{[100]}$ & 10.74 $(1.38)_{[100]}$ & 0.45 $(0.25)_{[100]}$ & 32.29 $(4.35)_{[100]}$ & 9.55 $(2.39)_{[100]}$ & 0.20 $(0.08)_{[100]}$ \\
\hline
\end{tabular}
\end{sidewaystable}

\begin{sidewaystable}
{\textwidth=3\linewidth \caption{\label{t:sim4} Simulation study: Beta distributed variables, differing between classes, $K=2$ groups. Average Adjusted Rand index $\times 100$ (with standard errors $\times 100$, in brackets) for different methods. Numbers in squared brackets indicate the valid cases out of 100 runs. \aftergroup\aftergroup\aftergroup
\aftergroup\aftergroup\aftergroup
\aftergroup
\leavevmode}}
\centering \footnotesize
\begin{tabular}{lrrrrrrrrr}
\hline
 & \multicolumn{9}{c}{$n=50$}    \\
  \hline
  & \multicolumn{3}{c}{$p=50$} & \multicolumn{3}{c}{$p=100$} & \multicolumn{3}{c}{$p=500$} \\
   \hline
 & $100\%$ & $50\%$ &    $10\%$ & $100\%$ & $50\%$  & $10\%$ & $100\%$ & $50\%$  & $10\%$ \\
CU & 41.03 $(4.27)_{[100]}$ & 30.75 $(3.97)_{[100]}$ & 5.49 $(1.30)_{[100]}$ & 47.35 $(4.37)_{[100]}$ & 36.86 $(4.30)_{[100]}$ & 9.08 $(1.90)_{[100]}$ & 59.23 $(4.27)_{[100]}$ & 49.15 $(4.52)_{[100]}$ & 16.58 $(2.84)_{[100]}$ \\
  VU & 40.96 $(4.20)_{[100]}$ & 30.45 $(3.97)_{[100]}$ & 5.95 $(1.47)_{[100]}$ & 49.14 $(4.36)_{[100]}$ & 38.90 $(4.32)_{[100]}$ & 9.47 $(1.99)_{[100]}$ & 61.84 $(4.16)_{[100]}$ & 51.28 $(4.53)_{[100]}$ & 21.34 $(3.04)_{[100]}$ \\
  CS & 40.58 $(4.24)_{[100]}$ & 30.32 $(3.93)_{[100]}$ & 4.71 $(1.24)_{[100]}$ & 47.34 $(4.35)_{[100]}$ & 36.83 $(4.31)_{[100]}$ & 8.56 $(1.79)_{[100]}$ & 59.19 $(4.28)_{[100]}$ & 49.15 $(4.52)_{[100]}$ & 16.00 $(2.71)_{[100]}$ \\
  VS & 38.99 $(4.15)_{[100]}$ & 29.14 $(3.96)_{[100]}$ & 3.95 $(0.95)_{[100]}$ & 45.00 $(4.38)_{[100]}$ & 36.64 $(4.31)_{[100]}$ & 8.79 $(1.83)_{[100]}$ & 51.81 $(4.25)_{[100]}$ & 42.85 $(4.54)_{[100]}$ & 16.30 $(2.85)_{[100]}$ \\
  Mixt-N & 55.88 $(4.65)_{[100]}$ & 36.41 $(4.28)_{[100]}$ & 7.99 $(1.70)_{[100]}$ & 55.77 $(4.77)_{[100]}$ & 39.89 $(4.50)_{[100]}$ & 14.79 $(2.64)_{[100]}$ & 51.07 $(4.76)_{[100]}$ & 47.60 $(4.76)_{[100]}$ & 28.34 $(3.94)_{[100]}$ \\
\hline
\end{tabular}
\end{sidewaystable}

\begin{sidewaystable}
{\textwidth=3\linewidth \caption{\label{t:sim5} Simulation study: Beta (fifth scenario) distributed variables, differing between classes, $K=2$ groups. Average Adjusted Rand index $\times 100$ (with standard errors $\times 100$, in brackets) for different methods. Numbers in squared brackets indicate the valid cases out of 100 runs. \aftergroup\aftergroup\aftergroup
\aftergroup\aftergroup\aftergroup
\aftergroup
\leavevmode}}
\centering \footnotesize
\begin{tabular}{lrrrrrrrrr}
\hline
 & \multicolumn{9}{c}{$n=50$}    \\
  \hline
  & \multicolumn{3}{c}{$p=50$} & \multicolumn{3}{c}{$p=100$} & \multicolumn{3}{c}{$p=500$} \\
   \hline
 & $100\%$ & $50\%$ &    $10\%$ & $100\%$ & $50\%$  & $10\%$ & $100\%$ & $50\%$  & $10\%$ \\
 CU & 18.83 $(2.35)_{[100]}$ & 2.68 $(0.59)_{[100]}$ & 0.05 $(0.29)_{[100]}$ & 32.42 $(3.52)_{[100]}$ & 3.60 $(0.78)_{[100]}$ & 0.07 $(0.26)_{[100]}$ & 41.36 $(4.92)_{[100]}$ & 1.94 $(0.49)_{[100]}$ & -0.08 $(0.28)_{[100]}$ \\
  VU & 26.05 $(2.43)_{[100]}$ & 3.84 $(0.69)_{[100]}$ & 0.44 $(0.27)_{[100]}$ & 38.30 $(3.78)_{[100]}$ & 4.54 $(0.96)_{[100]}$ & 0.42 $(0.31)_{[100]}$ & 51.52 $(5.01)_{[100]}$ & 2.89 $(0.77)_{[100]}$ & 0.01 $(0.25)_{[100]}$ \\
  CS & 28.72 $(2.87)_{[100]}$ & 5.56 $(0.88)_{[100]}$ & 0.08 $(0.27)_{[100]}$ & 41.90 $(3.68)_{[100]}$ & 5.92 $(1.22)_{[100]}$ & 0.26 $(0.26)_{[100]}$ & 43.43 $(4.85)_{[100]}$ & 2.93 $(0.79)_{[100]}$ & -0.09 $(0.28)_{[100]}$ \\
  VS & 33.85 $(2.96)_{[100]}$ & 9.04 $(1.50)_{[100]}$ & 0.15 $(0.28)_{[100]}$ & 46.89 $(3.65)_{[100]}$ & 11.80 $(1.65)_{[100]}$ & 0.82 $(0.41)_{[100]}$ & 59.44 $(4.38)_{[100]}$ & 11.85 $(2.07)_{[100]}$ & 0.84 $(0.47)_{[100]}$ \\
  Mixt-N & 18.45 $(2.08)_{[100]}$ & 4.83 $(0.81)_{[100]}$ & 0.47 $(0.29)_{[100]}$ & 41.21 $(3.22)_{[100]}$ & 4.17 $(0.94)_{[100]}$ & 0.57 $(0.32)_{[100]}$ & 94.84 $(0.75)_{[100]}$ & 34.46 $(2.06)_{[100]}$ & 0.83 $(0.34)_{[100]}$ \\
\hline
\end{tabular}
\end{sidewaystable}


\begin{sidewaystable}
{\textwidth=3\linewidth \caption{\label{t:sim3.1a}Simulation study: independent identically distributed symmetric variables, $K=3$ groups. Average Adjusted Rand index $\times 100$ (with standard errors $\times 100$, in brackets) for different methods. Numbers in squared brackets indicate the valid cases out of 100 runs.\aftergroup\aftergroup\aftergroup
\aftergroup\aftergroup\aftergroup
\aftergroup
\leavevmode}}
\centering \footnotesize
\begin{tabular}{lrrrrrrrrr}
\hline
 & \multicolumn{9}{c}{$n=50$}    \\
  \hline
  & \multicolumn{3}{c}{$p=50$} & \multicolumn{3}{c}{$p=100$} & \multicolumn{3}{c}{$p=500$} \\
   \hline
 & $100\%$ & $50\%$ &    $10\%$ & $100\%$ & $50\%$  & $10\%$ & $100\%$ & $50\%$  & $10\%$ \\
CU & 87.83 $(0.79)_{[100]}$ & 63.51 $(1.43)_{[100]}$ & 0.64 $(0.27)_{[100]}$ & 94.07 $(0.58)_{[100]}$ & 76.93 $(1.06)_{[100]}$ & 0.51 $(0.20)_{[100]}$ & 98.73 $(0.28)_{[100]}$ & 92.60 $(0.76)_{[100]}$ & 5.04 $(1.12)_{[100]}$ \\
  VU & 86.75 $(0.79)_{[100]}$ & 64.01 $(1.41)_{[100]}$ & 0.75 $(0.29)_{[100]}$ & 93.47 $(0.69)_{[100]}$ & 76.94 $(1.38)_{[100]}$ & 0.30 $(0.14)_{[100]}$ & 97.21 $(0.38)_{[100]}$ & 90.84 $(1.00)_{[100]}$ & 0.17 $(0.09)_{[100]}$ \\
  CS & 87.78 $(0.79)_{[100]}$ & 63.44 $(1.47)_{[100]}$ & 0.59 $(0.25)_{[100]}$ & 94.02 $(0.58)_{[100]}$ & 76.34 $(1.09)_{[100]}$ & 0.51 $(0.20)_{[100]}$ & 98.67 $(0.28)_{[100]}$ & 92.12 $(0.82)_{[100]}$ & 4.60 $(1.01)_{[100]}$ \\
  VS & 84.73 $(0.92)_{[100]}$ & 62.31 $(1.37)_{[100]}$ & 0.81 $(0.33)_{[100]}$ & 90.96 $(1.08)_{[100]}$ & 74.02 $(1.70)_{[100]}$ & 0.19 $(0.12)_{[100]}$ & 90.71 $(1.83)_{[100]}$ & 80.57 $(2.37)_{[100]}$ & 0.25 $(0.10)_{[100]}$ \\
  Mixt-N & 49.81 $(1.90)_{[100]}$ & 30.11 $(2.22)_{[100]}$ & 0.57 $(0.43)_{[100]}$ & 53.52 $(2.43)_{[100]}$ & 34.26 $(2.41)_{[100]}$ & 0.37 $(0.40)_{[100]}$ & 54.69 $(2.20)_{[100]}$ & 36.80 $(2.54)_{[100]}$ & -0.00 $(0.02)_{[100]}$ \\
\hline
\end{tabular}
\end{sidewaystable}

 \begin{sidewaystable}
{\textwidth=3\linewidth \caption{\label{t:sim3.1b}Simulation study: dependent identically distributed symmetric variables, $K=3$ groups.  Average Adjusted Rand index $\times 100$ (with standard errors $\times 100$, in brackets) for different methods. Numbers in squared brackets indicate the valid cases out of 100 runs.\aftergroup\aftergroup\aftergroup
\aftergroup\aftergroup\aftergroup
\aftergroup
\leavevmode}}
\centering \footnotesize
\begin{tabular}{lrrrrrrrrr}
\hline
 & \multicolumn{9}{c}{$n=50$}    \\
  \hline
  & \multicolumn{3}{c}{$p=50$} & \multicolumn{3}{c}{$p=100$} & \multicolumn{3}{c}{$p=500$} \\
   \hline
 & $100\%$ & $50\%$ &    $10\%$ & $100\%$ & $50\%$  & $10\%$ & $100\%$ & $50\%$  & $10\%$ \\
 CU & 89.76 $(0.70)_{[100]}$ & 65.36 $(1.50)_{[100]}$ & 0.64 $(0.28)_{[100]}$ & 93.80 $(0.63)_{[100]}$ & 76.36 $(1.10)_{[100]}$ & 0.27 $(0.10)_{[100]}$ & 99.21 $(0.20)_{[100]}$ & 91.97 $(0.90)_{[100]}$ & 5.90 $(1.19)_{[100]}$ \\
  VU & 87.35 $(0.83)_{[100]}$ & 63.13 $(1.48)_{[100]}$ & 0.97 $(0.34)_{[100]}$ & 92.82 $(0.63)_{[100]}$ & 74.88 $(1.36)_{[100]}$ & 0.22 $(0.14)_{[100]}$ & 97.35 $(0.39)_{[100]}$ & 91.69 $(1.07)_{[100]}$ & 0.26 $(0.10)_{[100]}$ \\
  CS & 89.76 $(0.72)_{[100]}$ & 64.59 $(1.51)_{[100]}$ & 0.49 $(0.16)_{[100]}$ & 93.74 $(0.63)_{[100]}$ & 75.73 $(1.09)_{[100]}$ & 0.25 $(0.10)_{[100]}$ & 99.15 $(0.21)_{[100]}$ & 91.68 $(0.95)_{[100]}$ & 5.21 $(1.04)_{[100]}$ \\
  VS & 84.63 $(0.92)_{[100]}$ & 59.38 $(1.47)_{[100]}$ & 0.88 $(0.26)_{[100]}$ & 90.77 $(0.96)_{[100]}$ & 70.40 $(1.61)_{[100]}$ & 0.31 $(0.18)_{[100]}$ & 90.21 $(1.99)_{[100]}$ & 81.96 $(2.27)_{[100]}$ & 0.16 $(0.11)_{[100]}$ \\
  Mixt-N & 47.49 $(2.07)_{[100]}$ & 25.13 $(2.24)_{[100]}$ & 0.43 $(0.33)_{[100]}$ & 51.78 $(2.45)_{[100]}$ & 33.82 $(2.46)_{[100]}$ & 0.01 $(0.06)_{[100]}$ & 54.67 $(2.21)_{[100]}$ & 37.72 $(2.59)_{[100]}$ & 0.02 $(0.03)_{[100]}$ \\
\hline
\end{tabular}
\end{sidewaystable}

\begin{sidewaystable}
{\textwidth=3\linewidth \caption{\label{t:sim3.2a}Simulation study: independent identically distributed asymmetric variables, $K=3$ groups. Average Adjusted Rand index $\times 100$ (with standard errors $\times 100$, in brackets) for different methods. Numbers in squared brackets indicate the valid cases out of 100 runs.\aftergroup\aftergroup\aftergroup
\aftergroup\aftergroup\aftergroup
\aftergroup
\leavevmode}}
\centering \footnotesize
\begin{tabular}{lrrrrrrrrr}
\hline
 & \multicolumn{9}{c}{$n=50$}    \\
  \hline
  & \multicolumn{3}{c}{$p=50$} & \multicolumn{3}{c}{$p=100$} & \multicolumn{3}{c}{$p=500$} \\
   \hline
 & $100\%$ & $50\%$ &    $10\%$ & $100\%$ & $50\%$  & $10\%$ & $100\%$ & $50\%$  & $10\%$ \\
CU & 64.41 $(1.88)_{[100]}$ & 41.51 $(0.54)_{[100]}$ & 7.32 $(0.55)_{[100]}$ & 80.29 $(2.09)_{[100]}$ & 48.52 $(0.48)_{[100]}$ & 5.28 $(0.62)_{[100]}$ & 91.50 $(1.71)_{[100]}$ & 54.43 $(0.06)_{[100]}$ & 1.93 $(0.45)_{[100]}$ \\
  VU & 85.73 $(1.23)_{[100]}$ & 41.99 $(1.27)_{[100]}$ & 1.25 $(0.27)_{[100]}$ & 97.85 $(0.43)_{[100]}$ & 52.59 $(1.03)_{[100]}$ & 0.21 $(0.09)_{[100]}$ & 100.00 $(0.00)_{[100]}$ & 57.25 $(1.11)_{[100]}$ & -0.01 $(0.00)_{[100]}$ \\
  CS & 95.99 $(1.35)_{[100]}$ & 45.51 $(0.74)_{[100]}$ & 16.17 $(1.22)_{[100]}$ & 99.55 $(0.45)_{[100]}$ & 49.96 $(0.49)_{[100]}$ & 13.02 $(1.53)_{[100]}$ & 99.09 $(0.64)_{[100]}$ & 54.54 $(0.04)_{[100]}$ & 3.10 $(0.86)_{[100]}$ \\
  VS & 93.72 $(1.48)_{[100]}$ & 54.28 $(2.10)_{[100]}$ & 1.75 $(0.31)_{[100]}$ & 94.27 $(1.81)_{[100]}$ & 62.89 $(2.63)_{[100]}$ & 0.70 $(0.23)_{[100]}$ & 91.32 $(2.33)_{[100]}$ & 58.97 $(2.35)_{[100]}$ & 0.31 $(0.17)_{[100]}$ \\
  Mixt-N & 7.40 $(1.50)_{[100]}$ & -0.04 $(0.06)_{[100]}$ & -0.08 $(0.05)_{[100]}$ & 17.45 $(2.02)_{[100]}$ & 0.07 $(0.06)_{[100]}$ & 0.01 $(0.06)_{[100]}$ & 44.87 $(1.15)_{[100]}$ & 21.02 $(2.08)_{[100]}$ & 0.01 $(0.03)_{[100]}$ \\
\hline
\end{tabular}
\end{sidewaystable}

\begin{sidewaystable}
{\textwidth=3\linewidth \caption{\label{t:sim3.2b}Simulation study: dependent identically distributed asymmetric variables, $K=3$ groups.  Average Adjusted Rand index $\times 100$ (with standard errors $\times 100$, in brackets) for different methods. Numbers in squared brackets indicate the valid cases out of 100 runs.\aftergroup\aftergroup\aftergroup
\aftergroup\aftergroup\aftergroup
\aftergroup
\leavevmode}}
\centering \footnotesize
\begin{tabular}{lrrrrrrrrr}
\hline
 & \multicolumn{9}{c}{$n=50$}    \\
  \hline
  & \multicolumn{3}{c}{$p=50$} & \multicolumn{3}{c}{$p=100$} & \multicolumn{3}{c}{$p=500$} \\
   \hline
 & $100\%$ & $50\%$ &    $10\%$ & $100\%$ & $50\%$  & $10\%$ & $100\%$ & $50\%$  & $10\%$ \\
CU & 64.94 $(1.63)_{[100]}$ & 41.66 $(0.59)_{[100]}$ & 5.82 $(0.58)_{[100]}$ & 75.38 $(2.00)_{[100]}$ & 47.61 $(0.56)_{[100]}$ & 5.23 $(0.51)_{[100]}$ & 91.59 $(1.70)_{[100]}$ & 54.54 $(0.04)_{[100]}$ & 1.76 $(0.37)_{[100]}$ \\
  VU & 84.61 $(1.28)_{[100]}$ & 36.36 $(0.95)_{[100]}$ & 0.83 $(0.21)_{[100]}$ & 96.40 $(0.62)_{[100]}$ & 46.65 $(1.18)_{[100]}$ & 0.36 $(0.16)_{[100]}$ & 100.00 $(0.00)_{[100]}$ & 59.24 $(1.02)_{[100]}$ & -0.01 $(0.01)_{[100]}$ \\
  CS & 95.10 $(1.46)_{[100]}$ & 45.08 $(0.51)_{[100]}$ & 9.07 $(1.07)_{[100]}$ & 98.20 $(0.89)_{[100]}$ & 49.36 $(0.45)_{[100]}$ & 11.81 $(1.37)_{[100]}$ & 97.72 $(1.00)_{[100]}$ & 54.51 $(0.04)_{[100]}$ & 2.69 $(0.60)_{[100]}$ \\
  VS & 92.91 $(1.58)_{[100]}$ & 46.72 $(1.88)_{[100]}$ & 1.13 $(0.27)_{[100]}$ & 92.62 $(1.86)_{[100]}$ & 60.72 $(2.55)_{[100]}$ & 0.66 $(0.19)_{[100]}$ & 90.79 $(2.29)_{[100]}$ & 57.40 $(2.24)_{[100]}$ & 0.47 $(0.22)_{[100]}$ \\
  Mixt-N & 4.95 $(1.14)_{[100]}$ & 0.19 $(0.14)_{[100]}$ & -0.00 $(0.10)_{[100]}$ & 13.12 $(1.86)_{[100]}$ & 0.08 $(0.12)_{[100]}$ & -0.04 $(0.04)_{[100]}$ & 42.27 $(1.43)_{[100]}$ & 11.66 $(1.86)_{[100]}$ & 0.04 $(0.05)_{[100]}$ \\
\hline
\end{tabular}
 \end{sidewaystable}

\begin{sidewaystable}
{\textwidth=3\linewidth \caption{\label{t:sim3.3a}Simulation study: independent not identically distributed variables, $K=3$ groups. Average Adjusted Rand index $\times 100$ (with standard errors $\times 100$, in brackets) for different methods. Numbers in squared brackets indicate the valid cases out of 100 runs.\aftergroup\aftergroup\aftergroup
\aftergroup\aftergroup\aftergroup
\aftergroup
\leavevmode}}
\centering \footnotesize
\begin{tabular}{lrrrrrrrrr}
\hline
 & \multicolumn{9}{c}{$n=50$}    \\
  \hline
  & \multicolumn{3}{c}{$p=50$} & \multicolumn{3}{c}{$p=100$} & \multicolumn{3}{c}{$p=500$} \\
   \hline
 & $100\%$ & $50\%$ &    $10\%$ & $100\%$ & $50\%$  & $10\%$ & $100\%$ & $50\%$  & $10\%$ \\
CU & 65.09 $(0.94)_{[100]}$ & 31.15 $(1.23)_{[100]}$ & 1.46 $(0.27)_{[100]}$ & 84.00 $(0.92)_{[100]}$ & 39.76 $(1.28)_{[100]}$ & 1.31 $(0.37)_{[100]}$ & 99.70 $(0.13)_{[100]}$ & 55.51 $(0.87)_{[100]}$ & 5.03 $(0.99)_{[100]}$ \\
  VU & 60.26 $(1.22)_{[100]}$ & 29.47 $(0.96)_{[100]}$ & 2.54 $(0.41)_{[100]}$ & 84.54 $(0.93)_{[100]}$ & 41.38 $(1.23)_{[100]}$ & 1.51 $(0.30)_{[100]}$ & 99.39 $(0.22)_{[100]}$ & 52.51 $(0.40)_{[100]}$ & 0.22 $(0.04)_{[100]}$ \\
  CS & 64.89 $(1.07)_{[100]}$ & 29.37 $(1.06)_{[100]}$ & 1.01 $(0.22)_{[100]}$ & 84.64 $(0.95)_{[100]}$ & 38.99 $(1.34)_{[100]}$ & 1.13 $(0.29)_{[100]}$ & 99.76 $(0.12)_{[100]}$ & 49.82 $(1.08)_{[100]}$ & 4.84 $(0.95)_{[100]}$ \\
  VS & 60.87 $(1.50)_{[100]}$ & 29.07 $(1.09)_{[100]}$ & 2.06 $(0.39)_{[100]}$ & 76.05 $(1.55)_{[100]}$ & 40.15 $(1.36)_{[100]}$ & 1.24 $(0.29)_{[100]}$ & 83.29 $(2.54)_{[100]}$ & 45.29 $(1.59)_{[100]}$ & 0.15 $(0.09)_{[100]}$ \\
  Mixt-N & 40.57 $(0.61)_{[100]}$ & 28.76 $(0.97)_{[100]}$ & 1.53 $(0.32)_{[100]}$ & 44.60 $(0.72)_{[100]}$ & 36.90 $(0.76)_{[100]}$ & 2.86 $(0.45)_{[100]}$ & 59.81 $(2.20)_{[100]}$ & 47.18 $(0.85)_{[100]}$ & 18.96 $(0.97)_{[100]}$ \\
\hline
\end{tabular}
 \end{sidewaystable}

\begin{sidewaystable}
{\textwidth=3\linewidth \caption{\label{t:sim3.3b}
Simulation study: dependent not identically distributed variables, $K=3$ groups. Average Adjusted Rand index $\times 100$ (with standard errors $\times 100$, in brackets) for different methods. Numbers in squared brackets indicate the valid cases out of 100 runs.\aftergroup\aftergroup\aftergroup
\aftergroup\aftergroup\aftergroup
\aftergroup
\leavevmode}}
\centering \footnotesize
\begin{tabular}{lrrrrrrrrr}
\hline
 & \multicolumn{9}{c}{$n=50$}    \\
  \hline
  & \multicolumn{3}{c}{$p=50$} & \multicolumn{3}{c}{$p=100$} & \multicolumn{3}{c}{$p=500$} \\
   \hline
 & $100\%$ & $50\%$ &    $10\%$ & $100\%$ & $50\%$  & $10\%$ & $100\%$ & $50\%$  & $10\%$ \\
CU & 64.51 $(1.12)_{[100]}$ & 30.05 $(1.20)_{[100]}$ & 1.34 $(0.26)_{[100]}$ & 80.69 $(1.02)_{[100]}$ & 37.04 $(1.24)_{[100]}$ & 1.51 $(0.41)_{[100]}$ & 99.70 $(0.13)_{[100]}$ & 55.67 $(0.80)_{[100]}$ & 4.94 $(0.94)_{[100]}$ \\
  VU & 58.79 $(1.23)_{[100]}$ & 26.76 $(1.15)_{[100]}$ & 2.64 $(0.37)_{[100]}$ & 82.92 $(0.99)_{[100]}$ & 41.48 $(1.17)_{[100]}$ & 1.49 $(0.33)_{[100]}$ & 99.39 $(0.23)_{[100]}$ & 53.56 $(0.36)_{[100]}$ & 0.22 $(0.04)_{[100]}$ \\
  CS & 64.12 $(1.21)_{[100]}$ & 29.16 $(1.09)_{[100]}$ & 0.92 $(0.21)_{[100]}$ & 82.21 $(1.08)_{[100]}$ & 34.56 $(1.23)_{[100]}$ & 1.19 $(0.34)_{[100]}$ & 99.82 $(0.10)_{[100]}$ & 50.13 $(1.16)_{[100]}$ & 4.64 $(0.88)_{[100]}$ \\
  VS & 58.03 $(1.36)_{[100]}$ & 26.63 $(1.02)_{[100]}$ & 2.40 $(0.38)_{[100]}$ & 75.31 $(1.56)_{[100]}$ & 38.59 $(1.41)_{[100]}$ & 1.42 $(0.39)_{[100]}$ & 85.41 $(2.21)_{[100]}$ & 43.73 $(1.69)_{[100]}$ & 0.38 $(0.14)_{[100]}$ \\
  Mixt-N & 37.93 $(0.63)_{[100]}$ & 25.23 $(1.07)_{[100]}$ & 1.96 $(0.39)_{[100]}$ & 41.15 $(0.82)_{[100]}$ & 35.63 $(0.78)_{[100]}$ & 2.84 $(0.47)_{[100]}$ & 49.21 $(1.26)_{[100]}$ & 45.68 $(0.68)_{[100]}$ & 18.69 $(1.11)_{[100]}$ \\
\hline
\end{tabular}
\end{sidewaystable}

\begin{sidewaystable}
{\textwidth=3\linewidth \caption{\label{t:sim3.4} Simulation study: Beta distributed variables, differing between classes, $K=3$ groups. Average Adjusted Rand index $\times 100$ (with standard errors $\times 100$, in brackets) for different methods. Numbers in squared brackets indicate the valid cases out of 100 runs. \aftergroup\aftergroup\aftergroup
\aftergroup\aftergroup\aftergroup
\aftergroup
\leavevmode}}
\centering \footnotesize
\begin{tabular}{lrrrrrrrrr}
\hline
 & \multicolumn{9}{c}{$n=50$}    \\
  \hline
  & \multicolumn{3}{c}{$p=50$} & \multicolumn{3}{c}{$p=100$} & \multicolumn{3}{c}{$p=500$} \\
   \hline
 & $100\%$ & $50\%$ &    $10\%$ & $100\%$ & $50\%$  & $10\%$ & $100\%$ & $50\%$  & $10\%$ \\
CU & 27.61 $(2.66)_{[100]}$ & 23.88 $(2.76)_{[100]}$ & 0.61 $(0.36)_{[100]}$ & 29.42 $(2.94)_{[100]}$ & 17.81 $(2.33)_{[100]}$ & 0.67 $(0.30)_{[100]}$ & 36.81 $(3.49)_{[100]}$ & 21.36 $(2.63)_{[100]}$ & 1.13 $(0.34)_{[100]}$ \\
  VU & 29.40 $(2.65)_{[100]}$ & 22.58 $(2.88)_{[100]}$ & 0.46 $(0.20)_{[100]}$ & 32.58 $(3.17)_{[100]}$ & 18.59 $(2.36)_{[100]}$ & 0.05 $(0.03)_{[100]}$ & 36.15 $(3.77)_{[100]}$ & 18.21 $(2.66)_{[100]}$ & 0.00 $(0.00)_{[100]}$ \\
  CS & 27.69 $(2.66)_{[100]}$ & 23.55 $(2.68)_{[100]}$ & 0.45 $(0.26)_{[100]}$ & 29.08 $(2.93)_{[100]}$ & 17.72 $(2.33)_{[100]}$ & 0.62 $(0.30)_{[100]}$ & 36.83 $(3.48)_{[100]}$ & 21.42 $(2.64)_{[100]}$ & 1.15 $(0.34)_{[100]}$ \\
  VS & 29.75 $(2.57)_{[100]}$ & 22.46 $(2.84)_{[100]}$ & 0.47 $(0.21)_{[100]}$ & 31.65 $(2.96)_{[100]}$ & 17.28 $(2.23)_{[100]}$ & 0.03 $(0.06)_{[100]}$ & 33.45 $(3.36)_{[100]}$ & 18.53 $(2.56)_{[100]}$ & 0.11 $(0.12)_{[100]}$ \\
  Mixt-N & 34.18 $(2.47)_{[100]}$ & 35.81 $(2.85)_{[100]}$ & 3.04 $(0.71)_{[100]}$ & 33.29 $(2.38)_{[100]}$ & 25.48 $(2.32)_{[100]}$ & 4.48 $(0.92)_{[100]}$ & 36.63 $(2.61)_{[100]}$ & 31.59 $(2.34)_{[100]}$ & 13.92 $(1.73)_{[100]}$ \\
\hline
\end{tabular}
\end{sidewaystable}

\begin{sidewaystable}
{\textwidth=3\linewidth \caption{\label{t:sim3.5} Simulation study: Beta (fifth scenario) distributed variables, differing between classes, $K=3$ groups. Average Adjusted Rand index $\times 100$ (with standard errors $\times 100$, in brackets) for different methods. Numbers in squared brackets indicate the valid cases out of 100 runs. \aftergroup\aftergroup\aftergroup
\aftergroup\aftergroup\aftergroup
\aftergroup
\leavevmode}}
\centering \footnotesize
\begin{tabular}{lrrrrrrrrr}
\hline
 & \multicolumn{9}{c}{$n=50$}    \\
  \hline
  & \multicolumn{3}{c}{$p=50$} & \multicolumn{3}{c}{$p=100$} & \multicolumn{3}{c}{$p=500$} \\
   \hline
 & $100\%$ & $50\%$ &    $10\%$ & $100\%$ & $50\%$  & $10\%$ & $100\%$ & $50\%$  & $10\%$ \\
CU & 21.55 $(1.41)_{[100]}$ & 4.08 $(0.51)_{[100]}$ & 0.08 $(0.15)_{[100]}$ & 19.92 $(1.53)_{[100]}$ & 3.09 $(0.62)_{[100]}$ & 0.27 $(0.15)_{[100]}$ & 15.43 $(2.29)_{[100]}$ & 19.25 $(2.26)_{[100]}$ & 0.72 $(0.21)_{[100]}$ \\
  VU & 32.22 $(1.51)_{[100]}$ & 6.42 $(0.71)_{[100]}$ & 0.49 $(0.26)_{[100]}$ & 40.34 $(1.67)_{[100]}$ & 4.45 $(0.75)_{[100]}$ & 0.15 $(0.11)_{[100]}$ & 51.55 $(1.23)_{[100]}$ & 0.15 $(0.03)_{[100]}$ & 0.00 $(0.00)_{[100]}$ \\
  CS & 27.80 $(1.39)_{[100]}$ & 4.93 $(0.62)_{[100]}$ & 0.27 $(0.18)_{[100]}$ & 26.21 $(1.43)_{[100]}$ & 3.42 $(0.69)_{[100]}$ & 0.30 $(0.14)_{[100]}$ & 17.71 $(2.37)_{[100]}$ & 21.03 $(2.40)_{[100]}$ & 0.67 $(0.19)_{[100]}$ \\
  VS & 35.94 $(1.46)_{[100]}$ & 10.41 $(0.93)_{[100]}$ & 0.89 $(0.30)_{[100]}$ & 44.30 $(1.36)_{[100]}$ & 9.21 $(1.29)_{[100]}$ & 0.48 $(0.20)_{[100]}$ & 49.47 $(0.98)_{[100]}$ & 1.68 $(0.53)_{[100]}$ & 0.40 $(0.28)_{[100]}$ \\
  Mixt-N & 25.27 $(1.53)_{[100]}$ & 4.44 $(0.53)_{[100]}$ & 0.41 $(0.30)_{[100]}$ & 44.46 $(1.47)_{[100]}$ & 7.49 $(0.81)_{[100]}$ & 0.56 $(0.34)_{[100]}$ & 88.32 $(0.94)_{[100]}$ & 36.91 $(1.60)_{[100]}$ & 1.06 $(0.34)_{[100]}$ \\
\hline
\end{tabular}
\end{sidewaystable}


\begin{sidewaystable}
{\textwidth=3\linewidth \caption{\label{t:sim5.1a}
Simulation study: independent identically distributed symmetric variables, $K=5$ groups. Average Adjusted Rand index $\times 100$ (with standard errors $\times 100$, in brackets) for different methods. Numbers in squared brackets indicate the valid cases out of 100 runs.\aftergroup\aftergroup\aftergroup
\aftergroup\aftergroup\aftergroup
\aftergroup
\leavevmode}}
\centering \footnotesize
\begin{tabular}{lrrrrrrrrr}
\hline
 & \multicolumn{9}{c}{$n=50$}    \\
  \hline
  & \multicolumn{3}{c}{$p=50$} & \multicolumn{3}{c}{$p=100$} & \multicolumn{3}{c}{$p=500$} \\
   \hline
 & $100\%$ & $50\%$ &    $10\%$ & $100\%$ & $50\%$  & $10\%$ & $100\%$ & $50\%$  & $10\%$ \\
CU & 72.51 $(1.05)_{[100]}$ & 68.03 $(0.86)_{[100]}$ & 19.67 $(0.82)_{[100]}$ & 79.03 $(1.03)_{[100]}$ & 80.11 $(0.71)_{[100]}$ & 25.43 $(0.72)_{[100]}$ & 82.27 $(1.15)_{[100]}$ & 95.03 $(0.50)_{[100]}$ & 32.35 $(0.83)_{[100]}$ \\
  VU & 77.16 $(1.08)_{[100]}$ & 67.17 $(1.15)_{[100]}$ & 19.09 $(0.85)_{[100]}$ & 83.20 $(0.97)_{[100]}$ & 76.30 $(0.87)_{[100]}$ & 22.35 $(0.78)_{[100]}$ & 89.52 $(0.85)_{[100]}$ & 86.57 $(0.63)_{[100]}$ & 29.42 $(0.81)_{[100]}$ \\
  CS & 71.85 $(1.02)_{[100]}$ & 67.92 $(0.85)_{[100]}$ & 18.51 $(0.85)_{[100]}$ & 78.58 $(1.03)_{[100]}$ & 80.18 $(0.71)_{[100]}$ & 23.53 $(0.77)_{[100]}$ & 81.49 $(1.11)_{[100]}$ & 95.03 $(0.50)_{[100]}$ & 31.39 $(0.86)_{[100]}$ \\
  VS & 75.15 $(1.13)_{[100]}$ & 64.00 $(1.25)_{[100]}$ & 15.59 $(0.89)_{[100]}$ & 81.40 $(1.13)_{[100]}$ & 73.65 $(1.26)_{[100]}$ & 19.56 $(0.80)_{[100]}$ & 81.42 $(1.42)_{[100]}$ & 79.04 $(1.63)_{[100]}$ & 24.39 $(1.10)_{[100]}$ \\
  Mixt-N & 55.21 $(1.09)_{[100]}$ & 42.07 $(1.04)_{[100]}$ & 17.46 $(1.26)_{[100]}$ & 58.70 $(1.16)_{[100]}$ & 47.91 $(1.13)_{[100]}$ & 22.21 $(1.22)_{[100]}$ & 59.57 $(1.31)_{[100]}$ & 47.84 $(1.15)_{[100]}$ & 23.33 $(1.44)_{[100]}$ \\
\hline
\end{tabular}
\end{sidewaystable}

 \begin{sidewaystable}
{\textwidth=3\linewidth \caption{\label{t:sim5.1b}
Simulation study: dependent identically distributed symmetric variables, $K=5$ groups.  Average Adjusted Rand index $\times 100$ (with standard errors $\times 100$, in brackets) for different methods. Numbers in squared brackets indicate the valid cases out of 100 runs.\aftergroup\aftergroup\aftergroup
\aftergroup\aftergroup\aftergroup
\aftergroup
\leavevmode}}
\centering \footnotesize
\begin{tabular}{lrrrrrrrrr}
\hline
 & \multicolumn{9}{c}{$n=50$}    \\
  \hline
  & \multicolumn{3}{c}{$p=50$} & \multicolumn{3}{c}{$p=100$} & \multicolumn{3}{c}{$p=500$} \\
   \hline
 & $100\%$ & $50\%$ &    $10\%$ & $100\%$ & $50\%$  & $10\%$ & $100\%$ & $50\%$  & $10\%$ \\
CU & 71.69 $(1.10)_{[100]}$ & 67.62 $(1.06)_{[100]}$ & 16.97 $(0.76)_{[100]}$ & 78.70 $(1.05)_{[100]}$ & 82.79 $(0.75)_{[100]}$ & 24.06 $(0.74)_{[100]}$ & 82.82 $(1.15)_{[100]}$ & 94.14 $(0.52)_{[100]}$ & 32.16 $(0.82)_{[100]}$ \\
  VU & 75.59 $(1.01)_{[100]}$ & 64.77 $(1.08)_{[100]}$ & 17.31 $(0.75)_{[100]}$ & 82.70 $(0.86)_{[100]}$ & 78.92 $(0.83)_{[100]}$ & 22.51 $(0.72)_{[100]}$ & 89.06 $(0.75)_{[100]}$ & 86.01 $(0.61)_{[100]}$ & 29.45 $(0.83)_{[100]}$ \\
  CS & 71.23 $(1.09)_{[100]}$ & 67.82 $(1.07)_{[100]}$ & 15.69 $(0.79)_{[100]}$ & 78.67 $(1.05)_{[100]}$ & 82.84 $(0.76)_{[100]}$ & 22.82 $(0.73)_{[100]}$ & 82.07 $(1.15)_{[100]}$ & 94.14 $(0.52)_{[100]}$ & 31.76 $(0.82)_{[100]}$ \\
  VS & 74.03 $(1.05)_{[100]}$ & 61.32 $(1.07)_{[100]}$ & 14.54 $(0.77)_{[100]}$ & 80.92 $(1.19)_{[100]}$ & 74.50 $(1.39)_{[100]}$ & 19.85 $(0.82)_{[100]}$ & 81.51 $(1.46)_{[100]}$ & 79.06 $(1.69)_{[100]}$ & 24.43 $(1.08)_{[100]}$ \\
  Mixt-N & 53.44 $(1.00)_{[100]}$ & 39.86 $(1.01)_{[100]}$ & 14.49 $(1.16)_{[100]}$ & 58.37 $(1.06)_{[100]}$ & 46.26 $(1.07)_{[100]}$ & 19.82 $(1.22)_{[100]}$ & 59.18 $(1.26)_{[100]}$ & 46.98 $(1.15)_{[100]}$ & 22.76 $(1.44)_{[100]}$ \\
\hline
\end{tabular}
\end{sidewaystable}

\begin{sidewaystable}
{\textwidth=3\linewidth \caption{\label{t:sim5.2a}
Simulation study: independent identically distributed asymmetric variables, $K=5$ groups. Average Adjusted Rand index $\times 100$ (with standard errors $\times 100$, in brackets) for different methods. Numbers in squared brackets indicate the valid cases out of 100 runs.\aftergroup\aftergroup\aftergroup
\aftergroup\aftergroup\aftergroup
\aftergroup
\leavevmode}}
\centering \footnotesize
\begin{tabular}{lrrrrrrrrr}
\hline
 & \multicolumn{9}{c}{$n=50$}    \\
  \hline
  & \multicolumn{3}{c}{$p=50$} & \multicolumn{3}{c}{$p=100$} & \multicolumn{3}{c}{$p=500$} \\
   \hline
 & $100\%$ & $50\%$ &    $10\%$ & $100\%$ & $50\%$  & $10\%$ & $100\%$ & $50\%$  & $10\%$ \\
CU & 62.99 $(0.40)_{[100]}$ & 45.45 $(0.44)_{[100]}$ & 13.86 $(0.64)_{[100]}$ & 68.08 $(0.36)_{[100]}$ & 47.77 $(0.47)_{[100]}$ & 22.64 $(0.51)_{[100]}$ & 73.88 $(0.17)_{[100]}$ & 57.53 $(0.65)_{[100]}$ & 34.89 $(0.89)_{[100]}$ \\
  VU & 62.21 $(0.81)_{[100]}$ & 43.13 $(0.79)_{[100]}$ & 9.21 $(0.65)_{[100]}$ & 69.10 $(0.75)_{[100]}$ & 58.05 $(0.83)_{[100]}$ & 14.98 $(0.71)_{[100]}$ & 77.37 $(0.36)_{[100]}$ & 80.98 $(0.50)_{[100]}$ & 26.48 $(0.47)_{[100]}$ \\
  CS & 65.47 $(0.43)_{[100]}$ & 51.74 $(0.70)_{[100]}$ & 19.62 $(0.70)_{[100]}$ & 69.42 $(0.35)_{[100]}$ & 56.54 $(0.88)_{[100]}$ & 27.39 $(0.45)_{[100]}$ & 73.96 $(0.17)_{[100]}$ & 69.72 $(0.83)_{[100]}$ & 36.79 $(0.87)_{[100]}$ \\
  VS & 66.34 $(1.38)_{[100]}$ & 46.57 $(0.98)_{[100]}$ & 8.66 $(0.72)_{[100]}$ & 73.74 $(1.28)_{[100]}$ & 59.18 $(1.35)_{[100]}$ & 15.01 $(0.93)_{[100]}$ & 73.94 $(1.60)_{[100]}$ & 68.45 $(2.23)_{[100]}$ & 19.60 $(1.06)_{[100]}$ \\
  Mixt-N & 29.16 $(0.59)_{[100]}$ & 22.58 $(0.63)_{[100]}$ & 1.01 $(0.30)_{[100]}$ & 36.42 $(0.79)_{[100]}$ & 26.23 $(0.43)_{[100]}$ & 2.87 $(0.40)_{[100]}$ & 50.55 $(0.49)_{[100]}$ & 39.73 $(0.71)_{[100]}$ & 15.24 $(0.67)_{[100]}$ \\
\hline
\end{tabular}
\end{sidewaystable}

\begin{sidewaystable}
{\textwidth=3\linewidth \caption{\label{t:sim5.2b}Simulation study: dependent identically distributed asymmetric variables, $K=5$ groups.  Average Adjusted Rand index $\times 100$ (with standard errors $\times 100$, in brackets) for different methods. Numbers in squared brackets indicate the valid cases out of 100 runs.\aftergroup\aftergroup\aftergroup
\aftergroup\aftergroup\aftergroup
\aftergroup
\leavevmode}}
\centering \footnotesize
\begin{tabular}{lrrrrrrrrr}
\hline
 & \multicolumn{9}{c}{$n=50$}    \\
  \hline
  & \multicolumn{3}{c}{$p=50$} & \multicolumn{3}{c}{$p=100$} & \multicolumn{3}{c}{$p=500$} \\
   \hline
 & $100\%$ & $50\%$ &    $10\%$ & $100\%$ & $50\%$  & $10\%$ & $100\%$ & $50\%$  & $10\%$ \\
CU & 61.52 $(0.50)_{[100]}$ & 45.46 $(0.46)_{[100]}$ & 11.95 $(0.59)_{[100]}$ & 68.47 $(0.34)_{[100]}$ & 48.28 $(0.50)_{[100]}$ & 22.56 $(0.56)_{[100]}$ & 74.34 $(0.22)_{[100]}$ & 57.93 $(0.64)_{[100]}$ & 34.08 $(0.83)_{[100]}$ \\
  VU & 60.40 $(0.80)_{[100]}$ & 42.27 $(0.80)_{[100]}$ & 6.29 $(0.54)_{[100]}$ & 71.26 $(0.79)_{[100]}$ & 58.98 $(0.90)_{[100]}$ & 12.69 $(0.83)_{[100]}$ & 77.82 $(0.56)_{[100]}$ & 80.95 $(0.51)_{[100]}$ & 26.33 $(0.41)_{[100]}$ \\
  CS & 63.67 $(0.51)_{[100]}$ & 50.99 $(0.79)_{[100]}$ & 14.93 $(0.73)_{[100]}$ & 69.58 $(0.32)_{[100]}$ & 56.28 $(0.91)_{[100]}$ & 27.33 $(0.44)_{[100]}$ & 74.38 $(0.22)_{[100]}$ & 70.06 $(0.82)_{[100]}$ & 36.30 $(0.79)_{[100]}$ \\
  VS & 64.45 $(1.21)_{[100]}$ & 44.71 $(0.99)_{[100]}$ & 5.46 $(0.58)_{[100]}$ & 76.40 $(1.66)_{[100]}$ & 60.53 $(1.52)_{[100]}$ & 12.05 $(0.88)_{[100]}$ & 73.53 $(1.62)_{[100]}$ & 68.75 $(2.19)_{[100]}$ & 17.99 $(1.11)_{[100]}$ \\
  Mixt-N & 27.41 $(0.61)_{[100]}$ & 19.66 $(0.74)_{[100]}$ & 1.10 $(0.27)_{[100]}$ & 33.55 $(0.81)_{[100]}$ & 24.30 $(0.59)_{[100]}$ & 1.61 $(0.39)_{[100]}$ & 47.20 $(0.61)_{[100]}$ & 35.95 $(0.59)_{[100]}$ & 11.92 $(0.71)_{[100]}$ \\
\hline
\end{tabular}
 \end{sidewaystable}

\begin{sidewaystable}
{\textwidth=3\linewidth \caption{\label{t:sim5.3a}
Simulation study: independent not identically distributed variables, $K=5$ groups. Average Adjusted Rand index $\times 100$ (with standard errors $\times 100$, in brackets) for different methods. Numbers in squared brackets indicate the valid cases out of 100 runs.\aftergroup\aftergroup\aftergroup
\aftergroup\aftergroup\aftergroup
\aftergroup
\leavevmode}}
\centering \footnotesize
\begin{tabular}{lrrrrrrrrr}
\hline
 & \multicolumn{9}{c}{$n=50$}    \\
  \hline
  & \multicolumn{3}{c}{$p=50$} & \multicolumn{3}{c}{$p=100$} & \multicolumn{3}{c}{$p=500$} \\
   \hline
 & $100\%$ & $50\%$ &    $10\%$ & $100\%$ & $50\%$  & $10\%$ & $100\%$ & $50\%$  & $10\%$ \\
CU & 50.42 $(0.53)_{[100]}$ & 43.75 $(0.67)_{[100]}$ & 16.51 $(0.58)_{[100]}$ & 61.89 $(0.54)_{[100]}$ & 58.27 $(0.86)_{[100]}$ & 24.92 $(0.66)_{[100]}$ & 70.74 $(0.27)_{[100]}$ & 81.85 $(0.77)_{[100]}$ & 43.95 $(0.69)_{[100]}$ \\
  VU & 51.27 $(0.86)_{[100]}$ & 37.84 $(0.78)_{[100]}$ & 14.48 $(0.66)_{[100]}$ & 60.67 $(0.84)_{[100]}$ & 56.19 $(1.10)_{[100]}$ & 19.62 $(0.68)_{[100]}$ & 69.37 $(0.77)_{[100]}$ & 84.22 $(0.70)_{[100]}$ & 32.47 $(0.75)_{[100]}$ \\
  CS & 52.95 $(0.53)_{[100]}$ & 45.61 $(0.73)_{[100]}$ & 16.71 $(0.54)_{[100]}$ & 63.67 $(0.57)_{[100]}$ & 59.23 $(0.81)_{[100]}$ & 24.52 $(0.69)_{[100]}$ & 70.97 $(0.25)_{[100]}$ & 81.85 $(0.77)_{[100]}$ & 43.77 $(0.66)_{[100]}$ \\
  VS & 57.36 $(1.08)_{[100]}$ & 39.26 $(0.95)_{[100]}$ & 11.95 $(0.53)_{[100]}$ & 62.10 $(1.37)_{[100]}$ & 52.82 $(1.10)_{[100]}$ & 15.82 $(0.70)_{[100]}$ & 68.21 $(1.20)_{[100]}$ & 72.91 $(1.76)_{[100]}$ & 24.86 $(1.11)_{[100]}$ \\
  Mixt-N & 37.86 $(0.79)_{[100]}$ & 29.14 $(0.65)_{[100]}$ & 7.33 $(0.55)_{[100]}$ & 42.71 $(0.85)_{[100]}$ & 33.73 $(0.78)_{[100]}$ & 10.75 $(0.63)_{[100]}$ & 26.72 $(0.89)_{[100]}$ & 27.92 $(0.81)_{[100]}$ & 15.54 $(0.83)_{[100]}$ \\
\hline
\end{tabular}
 \end{sidewaystable}

\begin{sidewaystable}
{\textwidth=3\linewidth \caption{\label{t:sim5.3b}Simulation study: dependent not identically distributed variables, $K=5$ groups. Average Adjusted Rand index $\times 100$ (with standard errors $\times 100$, in brackets) for different methods. Numbers in squared brackets indicate the valid cases out of 100 runs.\aftergroup\aftergroup\aftergroup
\aftergroup\aftergroup\aftergroup
\aftergroup
\leavevmode}}
\centering \footnotesize
\begin{tabular}{lrrrrrrrrr}
\hline
 & \multicolumn{9}{c}{$n=50$}    \\
  \hline
  & \multicolumn{3}{c}{$p=50$} & \multicolumn{3}{c}{$p=100$} & \multicolumn{3}{c}{$p=500$} \\
   \hline
 & $100\%$ & $50\%$ &    $10\%$ & $100\%$ & $50\%$  & $10\%$ & $100\%$ & $50\%$  & $10\%$ \\
CU & 50.04 $(0.60)_{[100]}$ & 43.16 $(0.67)_{[100]}$ & 16.51 $(0.55)_{[100]}$ & 60.77 $(0.59)_{[100]}$ & 58.29 $(0.83)_{[100]}$ & 23.28 $(0.60)_{[100]}$ & 70.74 $(0.19)_{[100]}$ & 83.65 $(0.77)_{[100]}$ & 43.42 $(0.65)_{[100]}$ \\
  VU & 49.59 $(0.69)_{[100]}$ & 38.95 $(0.79)_{[100]}$ & 14.40 $(0.59)_{[100]}$ & 62.02 $(0.85)_{[100]}$ & 57.11 $(1.01)_{[100]}$ & 18.10 $(0.64)_{[100]}$ & 69.64 $(0.72)_{[100]}$ & 83.81 $(0.75)_{[100]}$ & 32.24 $(0.73)_{[100]}$ \\
  CS & 52.68 $(0.57)_{[100]}$ & 45.47 $(0.71)_{[100]}$ & 16.31 $(0.62)_{[100]}$ & 62.59 $(0.57)_{[100]}$ & 59.57 $(0.79)_{[100]}$ & 23.78 $(0.61)_{[100]}$ & 70.83 $(0.17)_{[100]}$ & 83.65 $(0.77)_{[100]}$ & 43.04 $(0.66)_{[100]}$ \\
  VS & 54.63 $(0.99)_{[100]}$ & 40.62 $(0.87)_{[100]}$ & 12.66 $(0.59)_{[100]}$ & 64.56 $(1.30)_{[100]}$ & 54.42 $(1.28)_{[100]}$ & 15.57 $(0.72)_{[100]}$ & 68.46 $(1.20)_{[100]}$ & 71.40 $(1.90)_{[100]}$ & 24.41 $(1.07)_{[100]}$ \\
  Mixt-N & 37.31 $(0.78)_{[100]}$ & 29.06 $(0.73)_{[100]}$ & 7.27 $(0.66)_{[100]}$ & 42.94 $(0.89)_{[100]}$ & 32.41 $(0.71)_{[100]}$ & 10.42 $(0.66)_{[100]}$ & 27.21 $(0.90)_{[100]}$ & 26.82 $(0.93)_{[100]}$ & 17.92 $(0.85)_{[100]}$ \\
\hline
\end{tabular}
\end{sidewaystable}

\begin{sidewaystable}
{\textwidth=3\linewidth \caption{\label{t:sim5.4} Simulation study: Beta distributed variables, differing between classes, $K=5$ groups. Average Adjusted Rand index $\times 100$ (with standard errors $\times 100$, in brackets) for different methods. Numbers in squared brackets indicate the valid cases out of 100 runs. \aftergroup\aftergroup\aftergroup
\aftergroup\aftergroup\aftergroup
\aftergroup
\leavevmode}}
\centering \footnotesize
\begin{tabular}{lrrrrrrrrr}
\hline
 & \multicolumn{9}{c}{$n=50$}    \\
  \hline
  & \multicolumn{3}{c}{$p=50$} & \multicolumn{3}{c}{$p=100$} & \multicolumn{3}{c}{$p=500$} \\
   \hline
 & $100\%$ & $50\%$ &    $10\%$ & $100\%$ & $50\%$  & $10\%$ & $100\%$ & $50\%$  & $10\%$ \\
CU & 19.89 $(1.28)_{[100]}$ & 20.50 $(1.76)_{[100]}$ & 0.95 $(0.30)_{[100]}$ & 24.05 $(1.57)_{[100]}$ & 14.23 $(1.19)_{[100]}$ & 1.04 $(0.25)_{[100]}$ & 31.21 $(1.82)_{[100]}$ & 19.24 $(1.63)_{[100]}$ & 1.53 $(0.31)_{[100]}$ \\
  VU & 20.57 $(1.21)_{[100]}$ & 21.02 $(1.70)_{[100]}$ & 2.16 $(0.32)_{[100]}$ & 25.41 $(1.43)_{[100]}$ & 14.85 $(1.22)_{[100]}$ & 1.12 $(0.25)_{[100]}$ & 32.60 $(1.87)_{[100]}$ & 19.79 $(1.67)_{[100]}$ & 0.97 $(0.28)_{[100]}$ \\
  CS & 19.89 $(1.29)_{[100]}$ & 20.55 $(1.75)_{[100]}$ & 0.90 $(0.30)_{[100]}$ & 24.09 $(1.59)_{[100]}$ & 14.27 $(1.19)_{[100]}$ & 1.07 $(0.25)_{[100]}$ & 31.24 $(1.81)_{[100]}$ & 19.29 $(1.63)_{[100]}$ & 1.58 $(0.32)_{[100]}$ \\
  VS & 19.86 $(1.20)_{[100]}$ & 17.95 $(1.54)_{[100]}$ & 1.90 $(0.35)_{[100]}$ & 23.29 $(1.37)_{[100]}$ & 13.68 $(1.17)_{[100]}$ & 1.07 $(0.29)_{[100]}$ & 27.17 $(1.72)_{[100]}$ & 17.12 $(1.56)_{[100]}$ & 0.99 $(0.33)_{[100]}$ \\
  Mixt-N & 23.12 $(1.42)_{[100]}$ & 26.62 $(1.59)_{[100]}$ & 2.99 $(0.44)_{[100]}$ & 24.63 $(1.54)_{[100]}$ & 17.96 $(1.31)_{[100]}$ & 2.37 $(0.44)_{[100]}$ & 25.86 $(1.74)_{[100]}$ & 22.91 $(1.55)_{[100]}$ & 10.14 $(0.98)_{[100]}$ \\
\hline
\end{tabular}
\end{sidewaystable}

\begin{sidewaystable}
{\textwidth=3\linewidth \caption{\label{t:sim5.5} Simulation study: Beta (fifth scenario) distributed variables, differing between classes, $K=5$ groups. Average Adjusted Rand index $\times 100$ (with standard errors $\times 100$, in brackets) for different methods. Numbers in squared brackets indicate the valid cases out of 100 runs. \aftergroup\aftergroup\aftergroup
\aftergroup\aftergroup\aftergroup
\aftergroup
\leavevmode}}
\centering \footnotesize
\begin{tabular}{lrrrrrrrrr}
\hline
 & \multicolumn{9}{c}{$n=50$}    \\
  \hline
  & \multicolumn{3}{c}{$p=50$} & \multicolumn{3}{c}{$p=100$} & \multicolumn{3}{c}{$p=500$} \\
   \hline
 & $100\%$ & $50\%$ &    $10\%$ & $100\%$ & $50\%$  & $10\%$ & $100\%$ & $50\%$  & $10\%$ \\
CU & 34.29 $(0.17)_{[100]}$ & 30.59 $(0.28)_{[100]}$ & 28.77 $(0.38)_{[100]}$ & 34.77 $(0.12)_{[100]}$ & 31.71 $(0.24)_{[100]}$ & 31.76 $(0.24)_{[100]}$ & 35.53 $(0.00)_{[100]}$ & 33.55 $(0.22)_{[100]}$ & 34.04 $(0.22)_{[100]}$ \\
  VU & 33.13 $(0.34)_{[100]}$ & 30.85 $(0.34)_{[100]}$ & 30.81 $(0.36)_{[100]}$ & 33.46 $(0.27)_{[100]}$ & 32.24 $(0.26)_{[100]}$ & 33.32 $(0.24)_{[100]}$ & 34.94 $(0.26)_{[100]}$ & 35.16 $(0.12)_{[100]}$ & 23.71 $(1.67)_{[100]}$ \\
  CS & 34.32 $(0.18)_{[100]}$ & 30.59 $(0.29)_{[100]}$ & 28.68 $(0.48)_{[100]}$ & 34.77 $(0.12)_{[100]}$ & 31.73 $(0.24)_{[100]}$ & 31.80 $(0.25)_{[100]}$ & 35.53 $(0.00)_{[100]}$ & 33.55 $(0.22)_{[100]}$ & 34.04 $(0.22)_{[100]}$ \\
  VS & 33.12 $(0.39)_{[100]}$ & 30.79 $(0.34)_{[100]}$ & 28.35 $(0.73)_{[100]}$ & 33.65 $(0.29)_{[100]}$ & 31.76 $(0.32)_{[100]}$ & 30.16 $(0.80)_{[100]}$ & 34.72 $(0.34)_{[100]}$ & 34.07 $(0.30)_{[100]}$ & 20.14 $(1.67)_{[100]}$ \\
  Mixt-N & 32.08 $(0.60)_{[100]}$ & 28.17 $(0.46)_{[100]}$ & 27.11 $(0.42)_{[100]}$ & 34.06 $(0.71)_{[100]}$ & 27.75 $(0.44)_{[100]}$ & 26.64 $(0.45)_{[100]}$ & 52.19 $(1.16)_{[100]}$ & 32.98 $(0.64)_{[100]}$ & 26.83 $(0.41)_{[100]}$ \\
\hline
\end{tabular}
\end{sidewaystable}


\begin{sidewaystable}{\textwidth=3\linewidth \caption{\label{tab:SimTimes}
Average computation time (and the corresponding standard errors in brackets) of each clustering method for a single dataset from each scenario, 50\% of relevant variables, $K=2$, both dependent and independent variables.  \aftergroup\aftergroup\aftergroup
\aftergroup\aftergroup\aftergroup
\aftergroup\leavevmode}}
\begin{tabular}{{l}*{6}{r}}
\hline
 Method & $n=p=50$&  $n=50$, $p=500$& $n=500$, $p=50$ & $n=p=500$ & $n=5000$, $p=100$ & $n=50000$, $p=100$\\
\hline
CU & 0.41 (0.09) & 2.99 \phantom{11}(0.08) & 0.53 \phantom{1}(0.03) & 4.58 \phantom{1}(0.16) & 7.46 \phantom{11}(1.35) & 49.95 \phantom{11}(0.00) \\
  VU & 0.95 (0.22) & 6.95 \phantom{11}(1.83) & 1.96 \phantom{1}(0.65) & 28.43 (10.50) & 34.96 \phantom{1}(14.32) & 60.45 \phantom{11}(0.00) \\
  CS & 1.22 (0.22) & 15.98 \phantom{11}(2.58) & 1.57 \phantom{1}(0.15) & 12.59 \phantom{1}(0.87) & 15.87 \phantom{11}(4.73) & 68.82 \phantom{11}(0.00) \\
  VS & 3.14 (1.44) & 5.78\phantom{11} (0.12) & 4.40 \phantom{1}(1.98) & 13.82 \phantom{1}(3.48) & 56.32 \phantom{1}(18.41) & 120.38 \phantom{11}(0.00) \\
  Mixt-N & 0.24 (0.23) & 0.08 \phantom{11}(0.01) & 45.92 (24.07) & 0.82 \phantom{1}(0.02) & 114.55 \phantom{1}(64.09) & 394.68 \phantom{11}(0.00) \\
  MFA & 7.73 (2.13) & 459.02 (101.37) & 24.04  \phantom{1}(3.31) & 305.58 (36.11) & 317.48 \phantom{1}(34.72) & 2832.56 (367.84) \\
  $k$-means & 0.11 (0.10) & 0.05 \phantom{11}(0.01) & 0.05 \phantom{1}(0.00) & 0.55 \phantom{1}(0.01) & 1.19 \phantom{11}(0.02) & 12.05 \phantom{11}(0.00) \\
  PAM & 0.00 (0.00) & 0.00 \phantom{11}(0.00) & 0.02 \phantom{1}(0.00) & 0.10 \phantom{1}(0.00) & 2.99 \phantom{11}(0.07) & 486.63 \phantom{11}(0.00) \\
  h-avg & 0.00 (0.00) & 0.00 \phantom{11}(0.00) & 0.03 \phantom{1}(0.00) & 0.33 \phantom{1}(0.00) & 6.42 \phantom{11}(0.15) & NaN \phantom{111}(NA) \\
  SpeCl & 0.05 (0.02) & 0.24 \phantom{11}(0.20) & 3.04 \phantom{1}(0.17) & 2.90 \phantom{1}(0.26) & 2150.28 (153.00) & 2248.24 \phantom{11}(0.00) \\
  AffPr & 0.05 (0.02) & 0.05 \phantom{11}(0.01) & 3.14 \phantom{1}(0.36) & 3.06 \phantom{1}(0.37) & 879.92 \phantom{1}(98.54) & 489.64 \phantom{11}(0.00) \\
\hline
 \end{tabular}

 \end{sidewaystable}

\end{document}